\documentclass[%
 twocolumn,
 10pt,
superscriptaddress,
 amsmath,amssymb,
 aps,
 pra,
floatfix,
]{revtex4-2}

\usepackage{graphicx}
\usepackage{bm}
\usepackage{times}
\usepackage{amsmath,amssymb}
\usepackage{enumerate}
\usepackage{multirow}
\usepackage{qcircuit}
\usepackage{xcolor}

\usepackage{epstopdf}

\usepackage[colorlinks=true,linkcolor=blue,citecolor=red,linktocpage=true,breaklinks=true]{hyperref}

\newcommand{\eq}{\begin{equation}}
\newcommand{\en}{\end{equation}}
\newcommand{\eqa}{\begin{eqnarray}}
\newcommand{\ena}{\end{eqnarray}}
\newcommand{\tr}{\mathrm{Tr}}

\begin{document}

\title{Quantum Teleportation in Non-equilibrium Environments and Fixed-point Fidelity}

\author{Xiaokun \surname{Yan}}

\affiliation{College of Physics, Jilin University, Changchun 130022, China}
\affiliation{State Key Laboratory of Electroanalytical Chemistry, Changchun Institute of Applied Chemistry,
Chinese Academy of Sciences, Changchun 130022, China}

\author{Zhihai \surname{Wang}}

\affiliation{Center for Quantum Sciences and School of Physics, Northeast Normal University, Changchun 130024, China}

\author{Kun \surname{Zhang}}
\email{kunzhang@nwu.edu.cn}
\affiliation{School of Physics, Northwest University, Xi’an 710127, China}
\affiliation{Shaanxi Key Laboratory for Theoretical Physics Frontiers, Xi'an 710127, China}
\affiliation{Peng Huanwu Center for Fundamental Theory, Xi'an 710127, China}

\author{Jin \surname{Wang}}
\email{jin.wang.1@stonybrook.edu}
\affiliation{Department of Chemistry, Stony Brook University, and Department of Physics and Astronomy, Stony Brook University, Stony Brook, New York 11794, USA}

\date{\today}

\begin{abstract}

Quantum teleportation, a fundamental protocol in quantum information science, enables the transfer of quantum states through entangled particle pairs and classical communication channels. While ideal quantum teleportation requires maximally entangled states as resources, real-world implementations inevitably face environmental noise and decoherence effects. In this work, we investigate quantum teleportation in non-equilibrium environments with different temperatures or chemical potentials. We employ the Bloch-Redfield equation to characterize the non-equilibrium dynamics in both bosonic and fermionic reservoirs. Our analysis reveals that non-equilibrium conditions, namely the temperature difference and chemical potential difference, can significantly enhance the teleportation fidelity beyond what is achievable in equilibrium scenarios. Notably, under certain non-equilibrium conditions, the fidelities for all input states converge to the same value. We term this behavior the teleportation with a fixed-point fidelity. Importantly, at this fixed point, fidelity can be further enhanced by tuning the non-equilibrium parameters. These findings offer crucial insights for the implementation of quantum communication protocols in realistic environments, demonstrating that non-equilibrium conditions not only improve the fidelity but also present a promising avenue for simplifying practical quantum teleportation schemes.

\end{abstract}

\maketitle

\section{Introduction}

Quantum teleportation is a process of transferring the quantum state of a particle to another \cite{PhysRevLett.70.1895}. This captivating phenomenon of quantum theory has been extensively investigated \cite{pirandola2015advances,bowen2001teleportation,verstraete2003optimal,bouwmeester1997experimental,PhysRevLett.80.1121,furusawa1998unconditional,olmschenk2009quantum,yin2012quantum}. Both theoretical proposals and subsequent experimental demonstrations have shown that quantum teleportation has significant potential for applications in quantum computing \cite{gottesman1999demonstrating,PhysRevLett.114.220502}, quantum communication \cite{gisin2007quantum,pirandola2017fundamental}, and the development of quantum networks \cite{huang2020identification,WOS:001396450700003}.

Ideal quantum teleportation requires a maximally entangled state to ensure a unit fidelity, which represents the perfect overlap between the input and output states. However, in real-world conditions, quantum systems are open, implying an inevitable interaction with the environment. Pure states will evolve into mixed states due to the external noises, leading to a deterioration of the entanglement \cite{verstraete2002fidelity,jiang2019quantum}. In this scenario, quantum teleportation has a fidelity of less than one. Moreover, the fidelity for different transferred states varies. Note that the advantage of quantum teleportation can only manifest when the average fidelity exceeds $2/3$, which is the average fidelity without using quantum entanglement \cite{horodecki1996teleportation,popescu1994bell}. 

The effects of noisy environments on various quantum correlations have been extensively studied, such as the entanglement \cite{yu2004finite,carvalho2004decoherence,harlender2022transfer}, Bell non-locality \cite{horodecki1999general,zhang2021entanglement}, and quantum discord \cite{mazzola2010sudden,li2021quantum,radhakrishnan2020multipartite}. Although it is expected that the environmental noises would damage the teleportation fidelity in general, special cases may be beneficial. For example, local environmental effects can enable certain initial states, which do not have an advantage in being used for teleportation, to achieve a genuine quantum teleportation fidelity \cite{badziag2000local,bandyopadhyay2002origin}. The types of noise and the initial entanglement will affect the choice of form, which in turn impacts the fidelity \cite{jiang2019quantumlocal}. If the noisy quantum channel is modeled by a single Lindblad operator, the average fidelity is always greater than 2/3 \cite{oh2002fidelity}. Systems with Heisenberg coupling through noisy quantum channels have been analyzed using the Lindblad master equation. The interaction is beneficial in enhancing the teleportation fidelity \cite{xie2021steady}. Furthermore, the teleportation fidelity subjected to local structured reservoirs exhibits abrupt variations coinciding with entanglement sudden death, while displaying oscillations in non-Markovian reservoirs due to the memory effects \cite{man2012quantum}. Recently, a noise-mitigation mechanism was proposed for both the discrete- and continuous-variable quantum teleportation schemes via investigating the non-Markovian decoherence dynamics \cite{PhysRevA.110.012442}.

In practice, multiple systems may be subject to different environments with different temperatures or chemical potentials, called the non-equilibrium environments. Non-equilibrium conditions are commonly characterized by the temperature or the chemical potential difference, which can drive the flow of energy or matter within the systems. Studies have shown that entanglement and coherence can persist in non-equilibrium steady states, which are distinct from the equilibrium cases \cite{wu2011quantum,lambert2007nonequilibrium,quiroga2007nonequilibrium,sinaysky2008dynamics}. Under certain conditions, non-equilibrium environments can even enhance the steady-state entanglement \cite{wang2018steady,zhang2021entanglement}. Since teleportation fidelity is strongly related to entanglement, investigating the impact of non-equilibrium environments on quantum teleportation becomes particularly significant.

In the standard quantum teleportation protocol, the sender Alice and the receiver Bob are typically assumed to share the same environment. However, this assumption is not practical when considering the fact that Alice and Bob exist at different spatial locations. When studying the effects of environmental noise, it is necessary to consider both equilibrium and non-equilibrium scenarios, which have not been thoroughly studied in quantum teleportation. To the best of the authors' knowledge, only the continuous-variable teleportation under non-equilibrium environments has been studied, and the qubit case remains unexplored \cite{kitajima2014application}.


In this work, we investigate quantum teleportation in non-equilibrium environments using the quantum master equation framework. Specifically, we employ the Bloch-Redfield equation \cite{bloch1957generalized,redfield1957theory}, a non-secularized Markovian quantum master equation that has been widely applied in various studies of open quantum systems \cite{redfield1996relaxation,pollard1996advances,ishizaki2009adequacy,lee2015coherent,novoderezhkin2004coherent,WOS:000349847000005}. Compared to the Lindblad master equation, the Bloch-Redfield equation, which does not rely on the secular approximation, provides a more accurate description of non-equilibrium steady states \cite{zhang2014curl,li2015steady,huangfu2018steady,zhang2015landscape,zhang2015shape,wang2018coherence,guarnieri2018steady}. Furthermore, it captures non-equilibrium phenomena with greater precision, making it particularly suitable for our investigation. However, it is important to note that the Bloch-Redfield equation has certain limitations, notably its potential failure to guarantee the positivity of the density matrix \cite{ishizaki2009adequacy,jeske2015bloch,spohn1980kinetic}. However, the positivity of the density matrix can be restored without the secular approximation in some cases by restricting the initial conditions to those compatible with the system-bath correlations \cite{suarez1992memory}, or by using a consistent noise model for the bath \cite{jeske2015bloch}.  In our calculations, we carefully account for this limitation by selecting parameters that ensure the density matrix maintains positivity throughout the evolution.

We study how the non-equilibrium environment influences the teleportation fidelity in two setups. The first setup applies the non-equilibrium steady states as the teleportation resource. Although quantum teleportation over long distances can be achieved without direct interactions among qubits, the steady states formed in such non-interacting systems inherently lack entanglement, rendering them ineffective for quantum teleportation. When modeling teleportation within quantum circuits, it is essential to account for the interactions between Alice's and Bob's qubits, as these interactions are crucial for generating and maintaining the necessary entanglement. Consequently, in our study of steady-state teleportation, we explicitly introduce interactions between the qubits to accurately simulate the dynamics of quantum teleportation in realistic quantum circuit settings. The second setup focuses on the dynamic influence of the non-equilibrium environments. We include both the non-equilibrium bosonic and fermionic reservoirs, which are characterized by temperature and chemical potential differences respectively. Through systematic calculations of teleportation fidelity under different non-equilibrium conditions, we clarify the impact of these environments on the performance of quantum teleportation. Notably, we find that specific non-equilibrium environments can enhance teleportation fidelity. In general, the teleportation fidelity depends on the input states. However, we notice the existence of \textit{fixed-point fidelity}, which is independent of the input states. We show that non-equilibrium environments, both bosonic and fermionic cases, can improve the fixed-point fidelity. This has a practical advantage, since it guarantees a fixed value of fidelity for all input states. 




This paper is organized as follows. In Sec. \ref{s2}, we present the theoretical framework for quantum teleportation and the expression for teleportation fidelity. We present our model and derive the Redfield equation in Sec. \ref{sec:model_RE}. In Sec. \ref{s3}, we analyze the teleportation fidelity using steady states as the entangled quantum resource under various non-equilibrium conditions. Sec. \ref{s4} focuses on the dynamical influences of the non-equilibrium environments on the teleportation fidelity. Finally, conclusions are drawn in Sec. \ref{s6}.

\section{TELEPORTATION AND MASTER EQUATION} \label{s2}

First, we review the standard teleportation protocol for the qubit setup. Then we present various fidelity measures, including the average and maximum fidelity, which quantify the quality of the teleportation. 

\subsection{Teleportation protocol}

Initially, Alice (A) and Bob (B) share a Bell state in the teleportation protocol of qubits, which can be expressed as:
\begin{eqnarray}
    \left | \beta(00)  \right \rangle_{AB} &=&\frac{1}{\sqrt{2}}(\left | 00  \right \rangle_{AB} + \left | 11  \right \rangle_{AB} ), \nonumber \\
    \left | \beta(01)  \right \rangle_{AB} &=&\frac{1}{\sqrt{2}}(\left | 01  \right \rangle_{AB} + \left | 10  \right \rangle_{AB} ), \nonumber \\
    \left | \beta(10)  \right \rangle_{AB} &=&\frac{1}{\sqrt{2}}(\left | 00  \right \rangle_{AB} - \left | 11  \right \rangle_{AB} ), \nonumber \\
    \left | \beta(11)  \right \rangle_{AB} &=&\frac{1}{\sqrt{2}}(\left | 01  \right \rangle_{AB} - \left | 10  \right \rangle_{AB} ).
\end{eqnarray}
The arbitrary single-qubit state $\left | \psi \right \rangle_{\bar{A}}$ that Alice wishes to teleport to Bob is denoted as
\begin{equation} \label{astate}
     \left | \psi \right \rangle_{\bar{A}} =\cos\frac{\theta}{2} \left | 0  \right \rangle_{\bar{A}} + e^{-i\phi}\sin\frac{\theta}{2} \left | 1  \right \rangle_{\bar{A}},
\end{equation}
where $\theta \in [0,\pi]$ and $\phi \in [0,2\pi)$ define a point on the Bloch sphere.

Assuming the initial maximally entangled state shared by Alice and Bob is $ \left | \beta(mn)  \right \rangle_{AB},(m,n=0,1)$. The initial state can then be written as
\begin{eqnarray} \label{state AAB 1}
     \left | \psi \right \rangle_{\bar{A}AB} &=& \left | \psi \right \rangle_{\bar{A}} \left | \beta(mn)  \right \rangle_{AB}
\end{eqnarray}
According to the standard teleportation protocol \cite{PhysRevLett.70.1895}, Alice implements the Bell measurements on the qubits $\bar{A}$ and $A$, and projects onto one of the Bell states. Then Alice sends the measurement outcome to Bob through classical communication. When Alice's measurement yields $ \left | \beta(ij)  \right \rangle_{\bar{A}A},(i,j=0,1)$, from Eq. (\ref{state AAB 1}), the density matrix of Bob's qubit after measurement is
\begin{multline}
    \rho_{B~(mn,ij)}^{out} \\
    = \frac{1}{p_{ij}}\tr_{\bar{A}A}(\left | \beta(ij)  \right \rangle\langle \beta(ij)|_{\bar{A}A}\otimes I_{B} \left | \psi \right \rangle\langle\psi|_{\bar{A}AB}),
\end{multline}
where \(p_{ij}=\tr(\left | \beta(ij)  \right \rangle\langle \beta(ij)|_{\bar{A}A}\otimes I_{B} \left | \psi \right \rangle\langle\psi|_{\bar{A}AB})\).
Bob can recover the state $\left | \psi \right \rangle_{\bar{A}}$ by applying appropriate a unitary operator corrections to his qubit and the final density matrix of Bob's qubit is 
\begin{align}\label{rfin}
    \rho_{B~(mn,ij)}^{fin}&=U_{mn,ij} \rho_{B~(mn,ij)}^{out}U^{\dagger}_{mn,ij}.
\end{align}
Here the notation \(U_{mn,ij}=U(|\beta(mn)\rangle_{AB}),|\beta(ij)\rangle_{\bar{A}A})\) labels the recover gate of Bob and  all \(U_{mn,ij}\) are organized in Table. \ref{table}, where rows represent the initial shared Bell state \(\left | \beta(mn)  \right \rangle_{AB}\) between Alice and Bob, and columns correspond to Alice's measurement outcomes \(\left | \beta(ij)  \right \rangle_{\bar{A}A}\). 

\begin{table}  
\caption{Recover gates of Bob}  
\begin{tabular}{c|cccc}  \hline\hline
 &$\left | \beta(00)  \right \rangle_{\bar{A}A}$ &$\left | \beta(01)  \right \rangle_{\bar{A}A}$ &$\left | \beta(10)  \right \rangle_{\bar{A}A}$& $\left | \beta(11)  \right \rangle_{\bar{A}A}$  
 \\
\hline  
$\left | \beta(00)  \right \rangle_{AB}$& $I$ & $X$ & $Z$ & $XZ$ \\
$\left | \beta(01)  \right \rangle_{AB}$& $X$ & $I$ & $XZ$ & $Z$ \\
$\left | \beta(10)  \right \rangle_{AB}$& $Z$ & $XZ$ & $I$ & $X$ \\
$\left | \beta(11)  \right \rangle_{AB}$& $XZ$ & $Z$ & $X$ & $I$ \\   \hline\hline
\end{tabular}  
\footnotetext{The first column represents the Bell state between Alice and Bob. The first row represents the Bell measurement results of Alice. The \(I,X,Z\) gates correspond to operators with the form of identity matrix, Pauli-X matrix and Pauli-Z matrix, respectively}  
\label{table}  
\end{table}

\subsection{Teleportation fidelity}

The success of quantum teleportation can be quantified by the fidelity between Bob's final state \(\rho_B^{fin}\) and the transferred state \(|\psi\rangle_{\bar{A}}\), namely
\begin{equation}\label{Fmn}
    F_{mn}(\theta,\phi) =\sum_{i,j=0}^1 p_{ij}{}_{\bar{A}}\langle \psi  | \rho_{B~(mn,ij)}^{fin}\left | \psi \right \rangle_{\bar{A}},
\end{equation}
The average fidelity over all possible input states is given by
\begin{equation}\label{Fav}
    \bar{F}_{mn}=\frac{1}{4\pi}\int_0^\pi d\theta \int_0^{2\pi} d\phi F_{mn}(\theta,\phi)\sin\theta.
\end{equation}
Both the state-dependent fidelity $F(\theta,\phi)$ and the average fidelity $\bar{F}$ equals one for the ideal teleportation. However, in realistic scenarios, the average fidelity is always less than one due to environmental interactions, which cause pure states to evolve into mixed states. The unitary corrections after the Bell measurements cannot perfectly recover the initial state when using mixed entangled resources. Horodecki et al. derived the upper bound for the teleportation fidelity with an arbitrary state $\rho_{AB}$ shared by Alice and Bob \cite{horodecki1996teleportation}, expressed as
\begin{equation}\label{fmax}
    \bar{F}_{max}=\frac{1}{2}\left(1+\frac{1}{3}\tr\sqrt{T^{\dagger}T}\right),
\end{equation}
where the elements of matrix $T$ are given by $T_{mn}=\tr(\rho_{AB}(\sigma_n \otimes \sigma_m))$, and $\sigma_{n,m}$ are the Pauli matrices.

In our work, we consider the entangled resource $\rho_{AB}$ to be an X-state in the form
\begin{equation} 
    \rho_{AB}^{X}=\begin{pmatrix}
  a & 0 & 0 & \alpha e^{i\beta }\\
 0 & b & \delta  e^{i\epsilon  } & 0 \\
  0& \delta  e^{-i\epsilon  } & c & 0\\
  \alpha e^{-i\beta }& 0 & 0 & d
\end{pmatrix}\label{rhoX},
\end{equation}
where all parameters are real and satisfy the normalization condition $ a+b+c+d=1 $. The initial state for the teleportation can be written as 
\begin{align}
    \rho_{\bar{A}AB}= \left | \psi \right \rangle_{\bar{A}} \left \langle \psi \right |_{\bar{A}} \otimes \rho^X_{AB}.
\end{align}
By employing the aforementioned standard quantum teleportation protocol, the density matrix of Bob’s qubit after Alice's measurement is
\begin{align}\label{routx}
    \rho_B^{out}=\frac{1}{p_{ij}}\tr_{\bar{A}A}(\left | \beta(ij)  \right \rangle\langle \beta(ij)|_{\bar{A}A}\otimes I_{B} \rho_{\bar{A}AB}).
\end{align}
By substituting the expression of Eq. (\ref{routx}) into Eq. (\ref{rfin}), one can obtain four fidelities of X-state  under the standard teleportation protocol from Eq. (\ref{Fmn})
\begin{multline}
\label{F0010}
    F_{00(10)}(\theta,\phi)=\frac{1}{4}\bigg [ b+c+3(a+d)+(-)2\alpha\cos\beta \\
     +(-)2\delta\cos\epsilon\cos2\phi + \cos2\theta f_{00(10)}\bigg ],
\end{multline}
\begin{multline}
\label{F0111}
    F_{01(11)}(\theta,\phi)=\frac{1}{4}\bigg [ a+d+3(b+c)+(-)2\delta\cos\epsilon  \\
    +(-)2\alpha\cos\beta\cos2\phi+\cos2\theta f_{01(11)} \bigg ].
\end{multline}
with the coefficients
{\begin{multline}
\label{f_00_10}
    f_{00(10)}\\
    = a-b-c
    +d-(+)2\alpha\cos\beta-(+)2\delta\cos\epsilon\cos2\phi,
\end{multline}}
\begin{multline}
\label{f_01_11}
    f_{01(11)} \\
    = -a+b+c-d-(+)2\delta\cos\epsilon-(+)2\alpha\cos\beta\cos2\phi.
\end{multline}
The subscript \((10)\) and \((11)\) in the Eq. (\ref{F0010}) and (\ref{F0111}) represent expressions of \(F_{10}\) and \(F_{11}\) when the expressions adopt the signs in brackets, while the same definition applies in expression Eq. (\ref{f_00_10}) and (\ref{f_01_11}). Later we would see that $f_{00(10)}$ and $f_{01(11)}$ play significant roles on the fixed-point fidelity.

The average fidelities can be derived by substituting Eqs. (\ref{F0010}) and (\ref{F0111}) into Eq. (\ref{Fav}) 
\begin{align}
    \bar{F}_{00(10)}&=\frac{1}{3}(1+a+d+(-)2\alpha \cos\beta),\\
   \bar{F}_{01(11)}&=\frac{1}{3}(1+b+c+(-)2\delta \cos\epsilon).
\end{align}
The corresponding maximum fidelity defined in Eq.~(\ref{fmax}) has the expression
\begin{multline}\label{FmaxX}
    \bar{F}_{max}=\frac{1}{2}\bigg(1+\frac{1}{3}({|a-b-c+d|} \\
    +2\sqrt{(\alpha-\delta)^2}+2\sqrt{(\alpha+\delta)^2})\bigg).
\end{multline}
Note that one always has $\bar{F}_{max} \geq \bar F$.

\section{\label{sec:model_RE} Model and Redfield equation}

Although the teleported state $|\psi\rangle_{\bar A}$ can also be subjected to environmental noises, we omit the influence of the environment on $\bar{A}$ for simplicity. We only consider that the entanglement resource shared with Alice and Bob is subjected to environmental noise. First, we set up the non-equilibrium two-qubit model. Then we present the Bloch-Redfield equation, which characterizes the influence of the non-equilibrium environments on the two qubits. 

\subsection{Non-equilibrium qubit model}

We consider two coupled qubits, each embedded in its respective reservoir. It corresponds to Alice and Bob's qubits experiencing distinct environments. The total Hamiltonian of the system and environments comprises three parts
\begin{equation}
    H=H_{AB}+H_R+V,
\end{equation}
where $H_{AB}$ is the two-qubit Hamiltonian; $H_R$ is the free Hamiltonian of two reservoirs, and $V$ is the interaction between the qubit and its corresponding environment. For simplicity, we set $\hbar = k_B= 1$ in the following. 

In the standard quantum teleportation protocol, Alice and Bob's qubits are relatively independent and do not interact directly, especially when considering long-distance quantum teleportation. However, if we consider the teleportation in quantum circuits, it is natural to include the interaction between Alice and Bob's qubits. To thoroughly investigate how reservoirs influence teleportation, we discuss both cases, namely the two qubits with and without interactions. We set the two-qubit Hamiltonian as
\begin{align}
    H_{AB}=\frac{\varepsilon_A}{2}\sigma^{z}_A  +\frac{\varepsilon_B}{2}\sigma^{z}_B  + \frac{\lambda}{2}\left(\sigma^{+}_A \sigma^{-}_B + \sigma^{-}_A \sigma^{+}_B \right),
\end{align}
where $\varepsilon_A$ and $\varepsilon_B$ represent the energy-level spacings of Alice and Bob's qubits, respectively. The coupling strength is $\lambda$ which represents the dipole-dipole interaction of the two atoms \cite{PhysRevLett.89.207902} and $\sigma^{+}(\sigma^{-})$ is the Pauli raising (lowering) operator. The Hamiltonian \(H_{AB}\) describes two spins subjected to
a z-direction homogeneous magnetic field, having the spin XY interaction, when the qubits represented by spin \cite{PhysRevLett.83.4204}. The eigen-energies and corresponding eigen-states of the two-qubit Hamiltonian $H_{AB}$ are given by \cite{zhang2021entanglement,PhysRevA.83.052110} 
\begin{align}
    E_1&=-\bar{\varepsilon}, & \left | 1 \right \rangle &= \left | 00 \right \rangle_{AB},  \nonumber \\
    E_2&=-\Omega, & \left | 2 \right \rangle &=\cos\theta\left | 01 \right \rangle_{AB} -\sin\theta\left | 10 \right \rangle_{AB} ,\nonumber \\
    E_3&=\Omega , & \left | 3 \right \rangle &= \sin\theta\left | 01 \right \rangle_{AB}+\cos\theta\left | 10 \right \rangle_{AB},  \nonumber \\
    E_4&=\bar{\varepsilon}, & \left | 4 \right \rangle&= \left | 11 \right \rangle_{AB},\label{eg}
\end{align}
where $\bar\varepsilon$ is the average energy-level space, namely $\bar{\varepsilon}=(\varepsilon_A+\varepsilon_B)/2$; $\Omega=\sqrt{(\varepsilon_B-\varepsilon_A)^2+\lambda^2}$ is the Rabi frequency; $\theta \in (0,\pi/2)$ is the mixing angle defined by $\tan2\theta=-\lambda/(\varepsilon_B-\varepsilon_A)$. In the symmetric qubit case $\varepsilon_A=\varepsilon_B$, we have $\theta=\pi/4$. For the general case, $\theta=\arctan(-\lambda/(\varepsilon_B-\varepsilon_A))/2$ when $\varepsilon_A+\varepsilon_B<0$ or $\theta=\pi/2+ \arctan(-\lambda/(\varepsilon_B-\varepsilon_A))/2$ when $\varepsilon_A+\varepsilon_B>0$. 

The free Hamiltonian of the reservoirs $H_R$ is characterized as 
\begin{align}
    H_R=\sum_{k_A} \omega_{k_{A}} b_{k_{A}}^{\dagger}b_{k_{A}}+\sum_{k_B} \omega_{k_{B}} b_{k_{B}}^{\dagger}b_{k_{B}},
\end{align}
where $b_{k_{A}}(b_{k_{A}}^{\dagger})$ or $b_{k_{B}}(b_{k_{B}}^{\dagger})$ is the annihilation (creation) operator for the $k$-th mode with frequencies $\omega_{k_{A}}$ or $\omega_{k_{B}}$ of the reservoirs interacting with Alice's or Bob's qubit. The interaction Hamiltonian of qubit-reservoir interaction under the rotating wave approximation can be expressed as 
\begin{multline}
    V = \sum_{k_A} g_{k_{A}}(\omega_{k_{A}})\left(\sigma^{-}_{A} b_{k_{A}}^{\dagger}+\sigma^{+}_{A} b_{k_{A}}\right)  \\
     +\sum_{k_B} g_{k_{B}}(\omega_{k_{B}})\left(\sigma^{-}_{B} b_{k_{B}}^{\dagger}+\sigma^{+}_{B} b_{k_{B}}\right),
\end{multline}
where $g_{k_{A}}(\omega_{k_{A}})$ and $g_{k_{B}}(\omega_{k_{B}})$ are qubit-reservoir coupling strengths,  which are assumed to be real and are functions of the frequency $\omega_{k_{A,B}}$. In the eigenbasis of $H_{AB}$, interaction Hamiltonian $V$ can be re-expressed as 
\begin{multline}
    V=\sum_{k_A} g_{k_{A}}(\omega_{k_{A}})(\eta_A+\xi_A) b_{k_{A}}^{\dagger} \\
    + \sum_{k_B}g_{k_{B}}(\omega_{k_{B}})(\eta_B+\xi_B) b_{k_{B}}^{\dagger} +\text{h.c.},
\end{multline}
where $\eta_{A,B}, \xi_{A,B}$ are transition operators.

The order of the eigenstate depends on $\lambda$, which gives two groups of transition operators. 
\begin{itemize}
    \item If $\lambda<2\sqrt{\varepsilon_A\varepsilon_B}$, namely the relatively weak coupling phase, the order of energy levels is $E_{1}<E_{2}<E_{3}<E_{4}$. The corresponding transition operators (in the eigenbasis) have the form 
 \begin{align} 
 \eta _ { A } & = \sin \theta ( | 3 \rangle \langle 4 | - | 1 \rangle \langle 2 | ) , \nonumber\\ \eta _ { B } & = \cos \theta ( | 3 \rangle \langle 4 | + | 1 \rangle \langle 2 | ) , \nonumber\\ \xi _ { A } & = \cos \theta ( | 2 \rangle \langle 4 | + | 1 \rangle \langle 3 | ) , \nonumber\\ \xi _ { B } & = \sin \theta ( | 1 \rangle \langle 3 | - | 2 \rangle \langle 4 | ) ,
 \end{align}
and the transition frequencies are $\varepsilon_{\pm}=\bar{\varepsilon}\pm \Omega$. The transition frequencies \(\varepsilon_{-}\) correspond the transition from the state \(| 2 \rangle\) to \(| 1 \rangle\) and the state \(| 4 \rangle\) to \(| 3 \rangle\). The transition frequencies \(\varepsilon_{+}\) correspond the transition from the state \(| 4 \rangle\) to \(| 2 \rangle\) and the state \(| 3 \rangle\) to \(| 1 \rangle\). 

    \item If $\lambda>2\sqrt{\varepsilon_A\varepsilon_B}$, namely the relatively strong coupling phase, the order of energy levels is $E_{2}<E_{1}<E_{4}<E_{3}$. The corresponding transition operators have the form 
\begin{align} \eta _ { A } & = \sin \theta ( | 4 \rangle \langle 3 | - | 2 \rangle \langle 1 | ) , \nonumber\\ \eta _ { B } & = \cos \theta ( | 4 \rangle \langle 3 | + | 2 \rangle \langle 1 | ) , \nonumber\\ \xi _ { A } & = \cos \theta ( | 2 \rangle \langle 4 | + | 1 \rangle \langle 3 | ) , \nonumber\\ \xi _ { B } & = \sin \theta ( | 1 \rangle \langle 3 | - | 2 \rangle \langle 4 | ), \end{align}
and the transition frequencies are $\varepsilon_{\pm}=\Omega \pm \bar{\varepsilon} $. The transition frequencies \(\varepsilon_{-}\) correspond the transition from the state \(| 1 \rangle\) to \(| 2 \rangle\) and the state \(| 3 \rangle\) to \(| 4 \rangle\). The transition frequencies \(\varepsilon_{+}\) correspond the transition from the state \(| 2 \rangle\) to \(| 4 \rangle\) and the state \(| 1 \rangle\) to \(| 3 \rangle\).
\end{itemize}

\subsection{Bloch-Redfield equation}
 
The Born-Markov quantum master equation in the interaction picture reads \cite{PhysRev.105.1206,redfield1957theory}
\begin{equation}
    \frac{d \tilde{\rho}_{AB}}{dt}=-\int_{0}^\infty ds \tr_R\left[V(t),[V(t-s),\tilde{\rho}_{AB}\otimes\tilde{\rho}_R]\right],
\end{equation}
where $\tilde{\rho}_{AB}$ is the reduced density operator of the coupled two qubits in the interaction picture, and $\tilde{\rho}_{R}$ is the initial density operator of the reservoirs, assuming in its own equilibrium state. Here $\tr_R$ denotes the partial trace with respect to the degrees of freedom of the reservoirs. 

Going back to the Schr$\ddot{\mathrm{o}}$dinger picture, the Bloch-Redfield equation is given by
\begin{equation}
    \frac{d\rho_{AB}}{dt}=-i[H_{AB},\rho_{AB}]+\sum_{j=A,B} \mathcal { D }_j(\rho_{AB}), 
\end{equation}
where $\mathcal {D}_j(\rho_{AB})$ is the dissipator given by
\begin{multline}
    \mathcal { D } _ { j } (\rho_{AB}) \\
    = \alpha _ { j } ( \varepsilon _ { - } ) ( \eta _ { j } ^ { \dagger } \rho_{AB} \eta _ { j } + \eta _ { j } ^ { \dagger } \rho_{AB} \xi _ { j } - \eta _ { j } \eta _ { j } ^ { \dagger } \rho_{AB} - \xi _ { j } \eta _ { j } ^ { \dagger } \rho_{AB}) \\
    + \alpha _ { j } ( \varepsilon _ { + } ) ( \xi _ { j } ^ { \dagger } \rho_{AB} \xi _ { j }   + \eta _ { j }^ { \dagger }   \rho_{AB} \xi _ { j }- \xi _ { j }\xi _ { j } ^ { \dagger }  \rho_{AB} - \eta _ { j }\xi _ { j } ^ { \dagger }  \rho_{AB}) \\
    + \beta _ { j } ( \varepsilon _ { - } )  ( \eta _ { j }  \rho_{AB} \eta _ { j }^ { \dagger } + \eta _ { j }  \rho_{AB} \xi _ { j }^ { \dagger } - \eta _ { j } ^ { \dagger }\eta _ { j }  \rho_{AB} - \xi _ { j } ^ { \dagger }  \eta _ { j }\rho_{AB}) \\
    + \beta _ { j } ( \varepsilon _ { +} )( \xi _ { j }  \rho_{AB} \xi _ { j }^ { \dagger }   + \eta _ { j }   \rho_{AB} \xi _ { j }^ { \dagger }- \xi _ { j }^ { \dagger } \xi _ { j }  \rho_{AB} - \eta _ { j }^ { \dagger } \xi _ { j }  \rho_{AB}) \\
    + \text{h.c.}
\end{multline}
Here the coefficients $\alpha _ { j }(\omega)$ and  $\beta_ { j }(\omega)$ are the dissipation rates, given by
\begin{align}
    \alpha _ { j }(\omega)= & \gamma_j(\omega)n_j(\omega), \nonumber \\
    \beta_ { j }(\omega) = & \gamma_j(\omega)(1\pm n_j(\omega)),
\end{align}
where the coupling spectrum $\gamma_j(\omega)$ is
\begin{align}
    \gamma_j(\omega)=\pi \sum_{k_j}|g_{k_j}(\omega)|^2\delta(\omega-\omega_{k_j})
\end{align}
and $n_j(\omega)$ is the Bose-Einstein (minus sign) or the Fermi-Dirac (plus sign) distribution
\begin{align}
    n_j(\omega)=\frac{1}{e^{(\omega-\mu_j)/T_j}\mp1}.
\end{align}
For bosonic reservoirs the sign of \( \beta_ { j }(\omega)\) is plus, while it is minus for fermionic reservoirs. Parameters $T_j$ and $\mu_j$ are the temperatures and chemical potentials of $j$-th reservoir, respectively. For bosonic reservoirs, such as photon or phonon baths, the particle number is not conserved with a vanishing chemical potential. Since the Bloch-Redfield equation is based on the assumption that the interaction between the system and the environment is
weak, we can further assume that the coupling spectra with different frequencies are much less than the energy scale of the two qubits, namely $g_{k_j}(\varepsilon_+),g_{k_j}(\varepsilon_-) \ll \varepsilon_A,\varepsilon_B$. Therefore, it is reasonable to view $g_{k_j}$ as constants (independent of the transition frequencies $\varepsilon_{\pm}$). 

The steady state can be solved by reformulating the Bloch-Redfield equation in the Liouville space \cite{wang2018steady,wang2019nonequilibrium}. A notable feature of the steady state is that 
it has the structure of the X density matrix, defined in Eq. (\ref{rhoX}), either in the eigenbasis or the local basis. 

The behavior of quantum coherence exhibits distinct characteristics under equilibrium and non-equilibrium conditions. In equilibrium environments with $T_A=T_B$ and $\mu_A=\mu_B$, the dynamics of the population and coherence are decoupled, resulting in the absence of steady-state coherence in the energy basis. However, under non-equilibrium conditions, these spaces become coupled, leading to the persistence of coherence in the steady state \cite{quiroga2007nonequilibrium,PhysRevA.78.062301,hu2018steady}. Specifically, the coherence is among the states \(|2\rangle\) and \(|3\rangle\) in the weak coupling regime, or \(|1\rangle\) and \(|4\rangle\) in the strong coupling regime. The steady-state coherence also contributes to the entanglement \cite{zhang2021entanglement,wang2018steady}.

\section{Teleportation in non-equilibrium steady state}\label{s3}

In this section, we analyze quantum teleportation, in which the steady state of a two-qubit system coupled to non-equilibrium environments is applied as the initial entangled resource. We adopt the teleportation protocol for steady states, which has the highest average fidelity of \(\bar{F}_{mn}\). In this section the \(\bar{F}_{11}\) is always the highest average fidelity, and the recovering gates are \(U_{11,ij}\). In most cases, high entanglement leads to high fidelity. However, in some special situations, fidelity and entanglement do not correlate positively \cite{nandi2018two}. Nevertheless, non-equilibrium settings that can enhance entanglement are of great significance for improving teleportation fidelity. In our work, the relationship between fidelity and concurrence is positively correlated in most cases, and the corresponding concurrence under different settings mentioned in the article has been provided in the Appendix \ref{A}. In the following, we investigate both bosonic and fermionic reservoirs, examining how non-equilibrium conditions affect quantum teleportation.

\subsection{Bosonic reservoirs}

\begin{figure}[t!]
    \centering
    \includegraphics[width=0.5\textwidth]{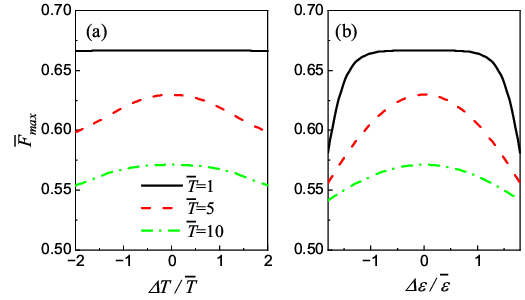}
    \caption{Maximum teleportation fidelity with the steady state of weak coupling system under the bosonic environments in terms of (a) the non-equilibrium condition \(\Delta T= T_A - T_B\) or (b) the detuned energy level $\Delta\varepsilon = \varepsilon_A-\varepsilon_B$. The average temperature is set as low with \(\bar{T}=1\) (black solid line), moderate with \(\bar{T}=5\) (red dashed line) or high with \(\bar{T}=10\) (green dashed dot line). The interaction strength is set as \(\lambda = 6\). Other parameters are set as \(\bar\varepsilon=10\) and \(g_A=g_B=0.05\).}
    \label{fig:l6_b_te}
\end{figure}

For bosonic reservoirs, we apply the temperature difference $\Delta T = T_A-T_B$ as the non-equilibrium condition. In the weak coupling phase $\lambda<\sqrt{\varepsilon_A\varepsilon_B}$, the limited steady state entanglement cannot support teleportation to have a quantum advantage (fidelity greater than 2/3) \cite{wang2018steady}. In the context of the weak coupling system, the ground state of the two-qubit system is a product state, which fundamentally limits the entanglement required for quantum teleportation. As a result, the teleportation fidelity at low average temperatures \(\bar T = (T_A+T_B)/2\) converges to the classical limit of $2/3$, as depicted by the black solid line in Fig. \ref{fig:l6_b_te} (a). The figure further illustrates that the fidelity exhibits a monotonous decline as the average temperature $\bar T$ increases. Additionally, both the temperature difference $|\Delta T|$ and the energy detuning $|\Delta\varepsilon|$ (Fig. \ref{fig:l6_b_te} (b)) contribute to a reduction in fidelity.

\begin{figure}[t!]
    \centering
    \includegraphics[width=0.5\textwidth]{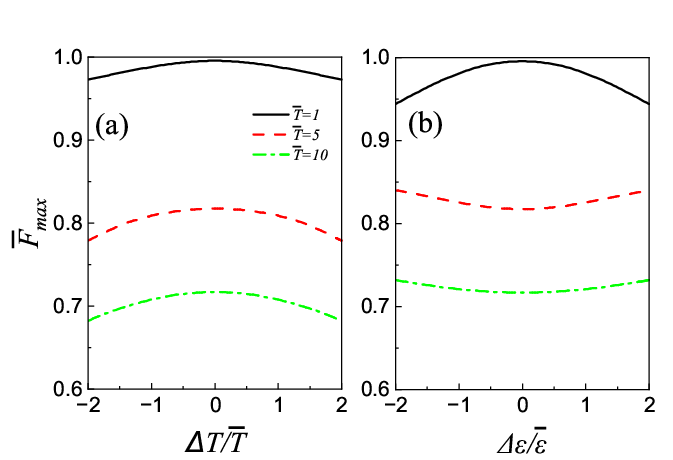}
    \caption{Maximum teleportation fidelity of the steady state under the bosonic environments in terms of (a) the non-equilibrium condition \(\Delta T= T_A - T_B\) or (b) the detuned energy level $\Delta\varepsilon = \varepsilon_A-\varepsilon_B$. The average temperature is set as low with \(\bar{T}=1\) (black solid line), moderate with \(\bar{T}=5\) (red dashed line) or high with \(\bar{T}=10\) (green dashed dot line). The interaction strength is set as \(\lambda = 30\). Other parameters are set as \(\varepsilon_A=\varepsilon_B=10\) and \(g_A=g_B=0.05\).}
    \label{fig:bssT}
\end{figure}

In the strong coupling phase $\lambda>\sqrt{\varepsilon_A\varepsilon_B}$, the maximum fidelity under different average temperatures $\bar T$ and temperature difference $\Delta T$ are shown in Fig. \ref{fig:bssT} (a). All the curves are decreasing with the increasing of \(|\Delta T|\). The temperature difference does not enhance the fidelity, but the overall fidelity value is notably higher than \(2/3\). This is because the strong coupling phase has an entangled ground state, while the excited states are product states. The higher temperature reservoir causes the corresponding qubit to be more easily excited. A common feature across different \(\bar{T}\) is that $\bar{F}_{max}$ decreases with increasing average temperature, which is consistent with the enhanced environmental decoherence effects. This universal behavior reflects the fundamental competition between quantum coherence and thermal noise.

We next examine the effect of energy detuning $\Delta\varepsilon=\varepsilon_A-\varepsilon_B$ on the teleportation fidelity. The detuning \(\Delta\varepsilon\) transforms eigenstates \(|2\rangle\) and \(|3\rangle\) from Bell states to partially entangled states, while also modifying the energy level spacing. In the strong coupling phase, the energy detuning \(\Delta\varepsilon\) enlarges the energy gap between the ground state and the first excited state, making the system harder to excite. Since the ground state is the entangled state \(|2\rangle\), this leads to enhanced entanglement. When the temperature is relatively low, the system stays at the ground state. Therefore, the detuning, which turns the maximally entangled ground state into a partially entangled state, leads to a decrease in fidelity as shown by the dark solid line in Fig. \ref{fig:bssT} (b). However, when the environmental temperature becomes elevated, the population of excited states increases, and the suppression of the excitation becomes predominant, consequently enhancing the system's entanglement and strengthening its fidelity. Fig. \ref{fig:bssT} (b) reveals the enhancement of $\bar{F}_{max}$ with average temperature \(\bar{T}=5\) and \(10\), as indicated by the red dashed line and the green dash-dot line. 

\begin{figure}[t!]
    \centering
    \includegraphics[width=0.53\textwidth]{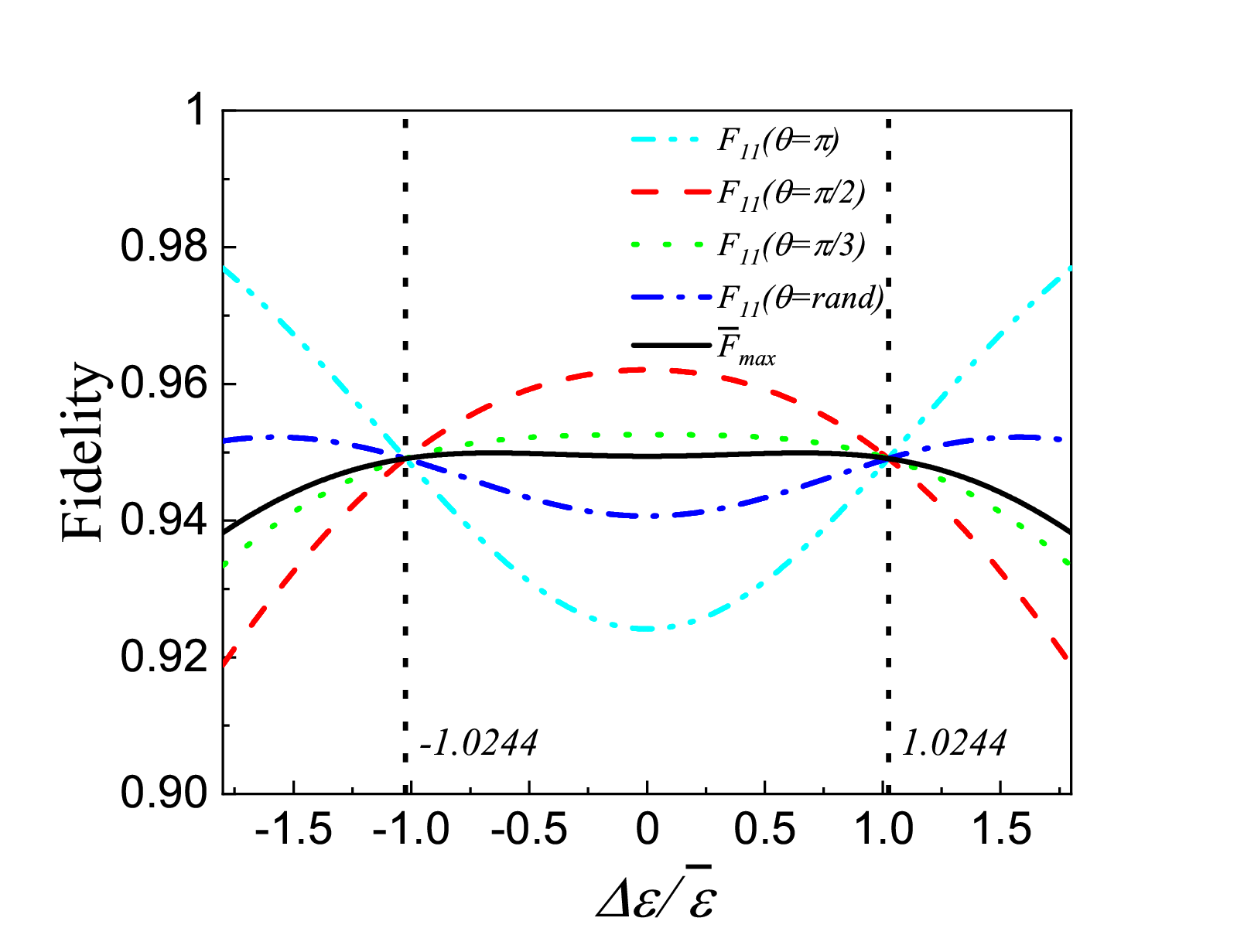}
    \caption{The fidelity of four teleported states and the maximal average fidelity \(\bar{F}_{max}\) (black solid line) in terms of the system's energy levels detuning \(\Delta \varepsilon=\varepsilon_A-\varepsilon_B\). The parameters \(\theta\) of four teleported states are \(\pi\) (cyan dashed dot dot), \(\pi/2\) (red dashed line),\(\pi/3\) (green dot line) and \(\theta=2.421\) (blue dashed dot line). Other parameter are set as \(T_A=T_B=2\), \(\varepsilon_A=\varepsilon_B=10\) and \(g_A=g_B=0.05\).}
    \label{fig:ssbdefp}
\end{figure}

Recall that $\bar{F}_{max}$ defined in Eq. (\ref{fmax}), is the maximum achievable average fidelity during teleportation. When teleporting specific states, the fidelities may vary from this average. A particular phenomenon emerges in the asymmetric two-qubit system where $\Delta\varepsilon\neq 0$, as shown by Fig. \ref{fig:ssbdefp}. The fidelities of all possible input states converge to the same value, matching the system's maximum fidelity $\bar{F}_{max}$ where $|\Delta\varepsilon/\bar{\varepsilon}|\approx 1.0244$. We call it the fixed-point fidelity. For a specific input state, the teleportation fidelity also depends on the measurement results. Suppose that Alice has the measurement results $|\beta(11)\rangle$. Then the fidelity $F_{11}(\theta,\phi)$ is given by Eq. (\ref{F0111}). The steady state of the non-equilibrium qubit model considered in our study has the form of the X-state defined in Eq. (\ref{rhoX}) with $\alpha = 0$. Therefore, we can see that $F_{11}(\theta,\phi)$ with $\alpha = 0$ is independent of the input state parameters $\theta$ and $\phi$ if the coefficient $f_{11}$ given in Eq. (\ref{f_01_11}) is zero. Although the conditions \(\alpha=0\) and \(f_{11}=0\) make \(F_{11}\) independent of \(\theta\) and \(\phi\), the fixed-point value typically remains below \(\bar{F}_{max}\). However, when the state parameter satisfies \(\epsilon=\pi\), \(F_{11}\) achieves the maximal value with \(F_{11}=\bar{F}_{max}\).

Fig. \ref{fig:ssbdefp} illustrates the fixed-point fidelity phenomenon. It shows how fidelity varies with energy detuning for four different transmitted states defined by Eq. (\ref{astate}) with $\theta=\pi$ (cyan dashed dot-dot line), $\theta=\pi/2$ (red dashed line), $\theta=\pi/3$ (green dot line), and $\theta=2.421$ (blue dashed dot line, this value is randomly chosen), each with random phases, $\phi$. The fidelity curves intersect at $|\Delta\varepsilon/\bar{\varepsilon}|\approx 1.0244$, indicating that $f_{11}=0$. At these intersection points, we observe that $\epsilon=\pi$ and $F_{11}=\bar{F}_{max}$ from the numerical results. 

\begin{figure}[t!]
    \centering
    \includegraphics[width=0.5\textwidth]{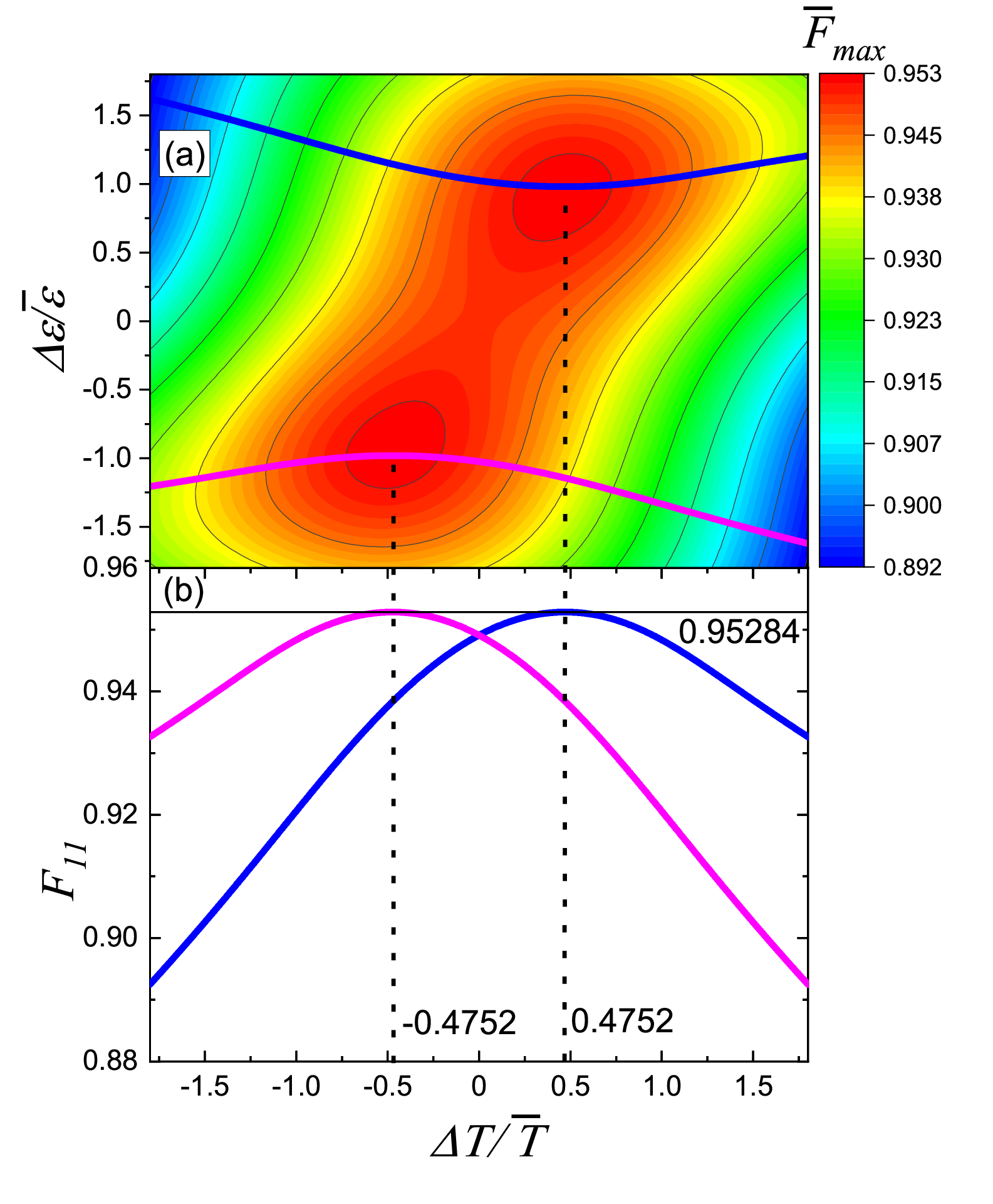}
    \caption{(a) The color-filled contour map shows the maximal fidelity in terms of the asymmetric energy detuning \(\Delta \varepsilon=\varepsilon_A-\varepsilon_B\) and the non-equilibrium environment with temperature difference \(\Delta T=T_A-T_B\). The two lines in (a) stand for the fixed-point fidelity and the corresponding fidelities \(\bar{F}_{11}\) are shown in (b). The horizontal line at 0.95284 marks the peak of \(\bar{F}_{11}\), while the vertical dashed line represents the corresponding \(\Delta T\) value \(\pm0.4752\). The parameters are set as \(T_A=T_B=2\), \(\varepsilon_A=\varepsilon_B=10\), \(\lambda=30\) and \(g_A=g_B=0.05\).}
    \label{fig:ssbte1}
\end{figure}

When both the non-equilibrium environment and energy level detuning coexist, their combined effects on $\bar{F}_{max}$ have intriguing patterns, as shown in Fig. \ref{fig:ssbte1} (a). The fidelity exhibits central symmetry, with maximum values occurring at $|\Delta T/\bar{T}| \approx 0.5$ and $|\Delta \varepsilon/\bar{\varepsilon}| \approx 0.9$. Most notably, the non-equilibrium conditions enhance fidelity with detuned two qubits beyond the equilibrium cases. The teleportation demonstrates better performance when the energy-temperature configuration present. Specifically, quantum teleportation fidelity peaks when Alice's qubit has a higher (lower) energy level, while interacting with a higher (lower) temperature reservoir. Considering that the ground state \(|2\rangle\) is an entangled state, higher energy levels can compensate for the excitation, thereby achieving stronger entanglement and enhanced fidelity. 

The fixed-point fidelity also emerges in Fig. \ref{fig:ssbte1} (a). Two distinct lines correspond to the condition $f_{11} = 0$. The numerical results of the density matrix $\rho_{AB}$ show that $\rho_{AB}(1,4)=\rho_{AB}^*(4,1)=0$. According to Eq. (\ref{F0111}), it effectively eliminates the phase dependence of $|\psi\rangle_{\bar{A}}$ on fidelity, leading to uniform teleportation performance across different input states. Furthermore, numerical analysis also shows that the difference between $F_{11}$ and $\bar{F}_{max}$ on those two lines is lower than $10^{-8}$ which means that $\epsilon\approx\pi$ and the numerical results indeed show $\epsilon\approx\pi$ to be the case.

As demonstrated in Fig. \ref{fig:ssbte1} (b), the fidelity $F_{11}$ along the fixed-point lines reaches the maximum value of 0.95284, which is remarkably close to the global maximum of 0.95287 in Fig. \ref{fig:ssbte1} (a). It suggests that the non-equilibrium condition can allow the steady state simultaneously achieve both the optimal fidelity and the fixed point fidelity. 

\subsection{Fermionic reservoirs}

In this section, we examine the quantum teleportation by two qubits coupled with non-equilibrium fermion reservoirs, where the non-equilibrium conditions are characterized by both the temperature and chemical potential differences. In the relatively weak coupling phase, the tunnel coupling strength is relatively weak compared to the  energy as in quantum dots system \cite{oosterkamp1998microwave}. The two qubits have the ground state \(|1\rangle\) with a particle number of zero. Fermionic reservoirs can excite the ground state into entanglement states \(|2\rangle\) and \(|3\rangle\). The highest excited state \(|4\rangle\) is also a product state with total particle number two. The effects of temperature and chemical potential on fidelity need to be discussed separately.

\begin{figure}[t!]
    \centering
    \includegraphics[width=0.5\textwidth]{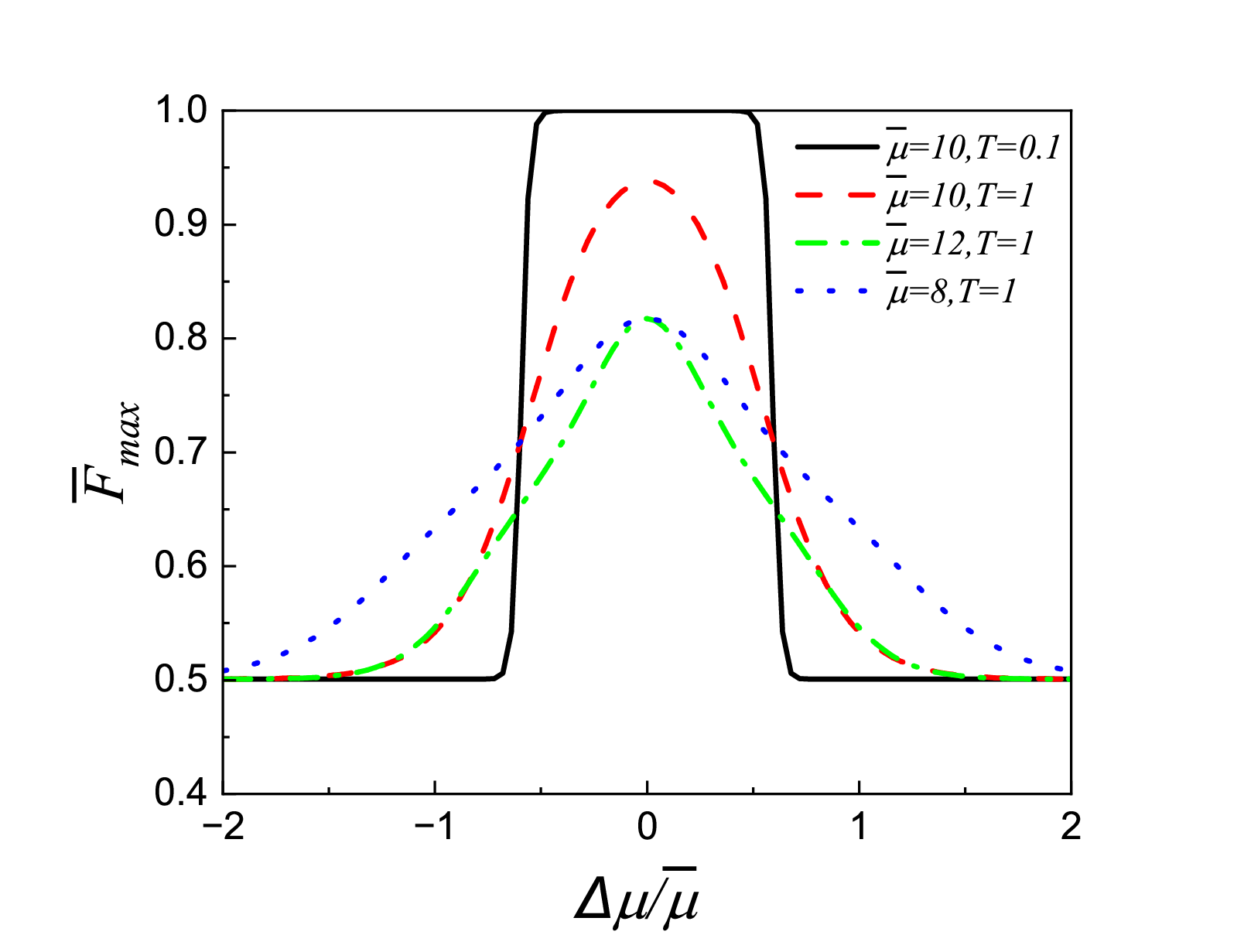}
    \caption{The maximal average fidelity in terms the non-equilibrium chemical potential \(\Delta \mu=\mu_A-\mu_B\). The average chemical potential is set as \(\bar{\mu}=10\) (black solid line for \(T=0.1\), red dashed line for \(T=1\)), \(\bar{\mu}=12\) (green dash-dot line for \(T=1\)) and \(\bar{\mu}=8\) (blued dot line for \(T=1\)). Other parameters are set as \(\varepsilon_A=\varepsilon_B=10\), and \(g_A=g_B=0.05\).}
    \label{fig:fdmu}
\end{figure}

Fig. \ref{fig:fdmu} illustrates how fidelity varies under chemical potential imbalance \(\Delta \mu=\mu_A-\mu_B\) when temperatures are equal across reservoirs. The Fermionic reservoir excites the system by means of the change in particle number. The resonant point \(\mu=\bar{\varepsilon}\) maximizes the population of the single-electron excited states \(|2\rangle\) and \(|3\rangle\). At relatively low temperature (\(T=0.1\)), which is shown as the black solid line in Fig. \ref{fig:fdmu}, the two qubits can have the fidelity \(\bar{F}_{max}\) approximately equal to 1 around the equilibrium regime. However, as \(|\Delta\mu/\bar{\mu}|\) exceeds a certain value (in this case the value is about 0.6), the fidelity \(\bar{F}_{max}\) exhibits rapid decay to 0.5 and the advantage of teleportation is destroyed. When the temperature rises to 1, the non-equilibrium condition \(\Delta\mu/\bar{\mu}\) reduces \(\bar{F}_{max}\) as shown the red dashed line in Fig. \ref{fig:fdmu}. In the case that \(\bar{\mu}\) deviates from the resonant point, the maximum \(\bar{F}_{max}\) is lower than that of \(\mu=\bar{\varepsilon}\) as indicated by the green dash-dot line (\(\bar{\mu}=12\)) and blue dotted line (\(\bar{\mu}=8\)) in Fig. \ref{fig:fdmu}.

\begin{figure}[t!]\label{fig:fdt}
    \centering
    \includegraphics[width=0.5\textwidth]{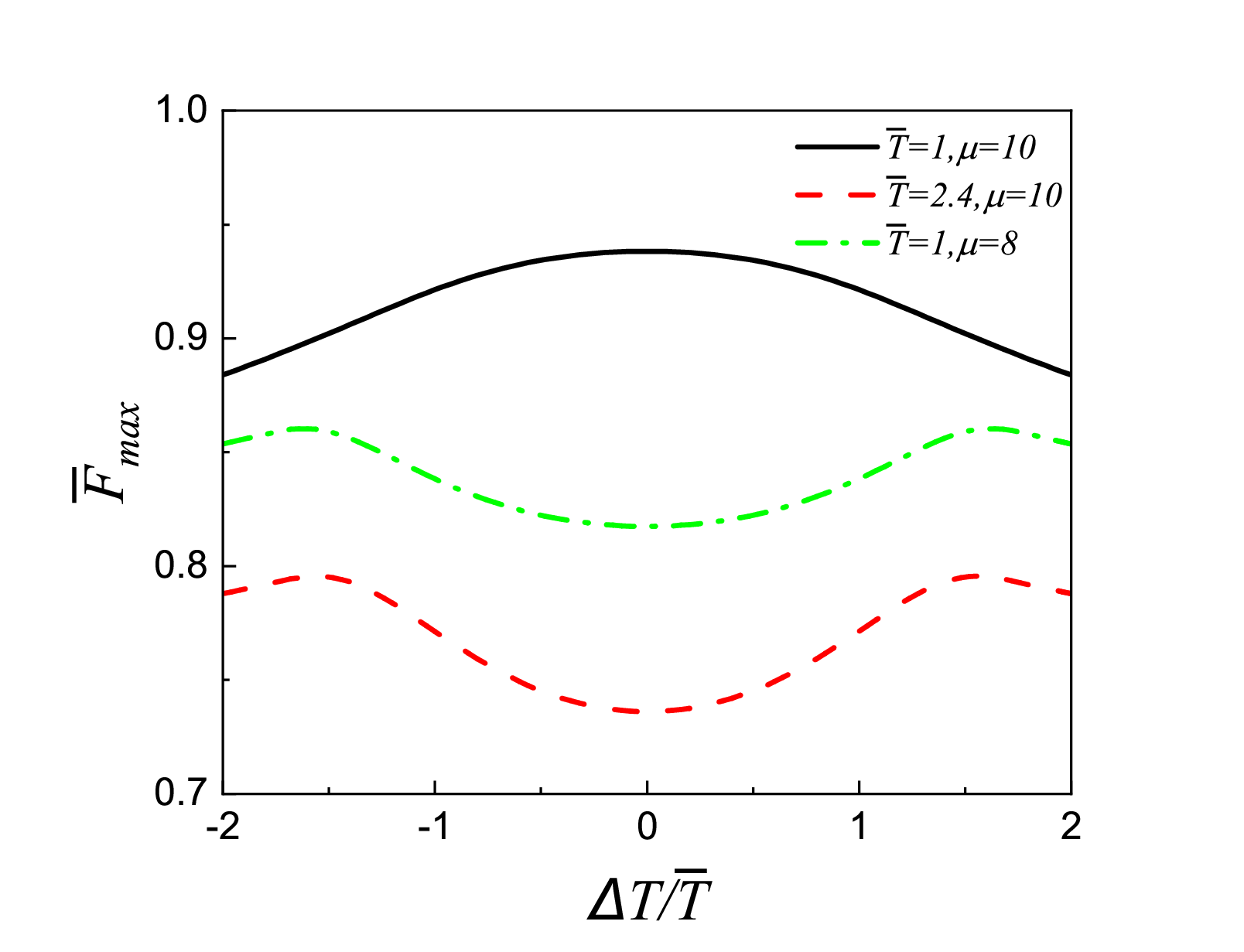}
    \caption{The maximal average fidelity in terms of the non-equilibrium temperature difference \(\Delta T=T_A-T_B\). The average temperature is set as \(\bar{T}=1\) (black solid line for \(\mu=10\)), \(\bar{T}=2.4\) (red dashed line for \(\mu=10\)), and \(\bar{T}=10\) (green dashed dot line for \(\mu=8\)). The interaction strength is \(\lambda=6 \). Other parameters are set as \(\varepsilon_A=\varepsilon_B=10\) and \(g_A=g_B=0.05\).}  
\end{figure}

The impact of temperature differences (from the fermionic reservoir) on teleportation fidelity is shown in Fig. \ref{fig:fdt}. At the resonant point \(\mu_{A,B}=\bar{\varepsilon}\), the black solid line shows that temperature difference \(\Delta T/\bar{T}\) reduces the fidelity when the average temperature is low. \(\Delta T/\bar{T}\) can enhance \(\bar{F}_{max}\) by increasing the average temperature (red dashed line) or the chemical potential deviating from the resonant point (green dash-dot line). For the red dashed line, where the average temperature \(\bar{T}=2.4\), the excitation effects of both temperature and chemical potential are stronger, which increases the population of state \(|4\rangle\) and impairs the entanglement of system leading to lower \(\bar{F}_{max}\). The presence of temperature difference \(\Delta T\) can decrease the population of state \(|4\rangle\) by diminishing the excitation of the qubit connecting with lower temperature and increases the entanglement of system. For the green dash-dot line, the average chemical potential \(\bar{\mu}=8\), the effect of \(\Delta T\) still decreases the population of state \(|4\rangle\), so the entanglement of system still increased. 

\begin{figure}[t!]
    \centering
    \includegraphics[width=0.5\textwidth]{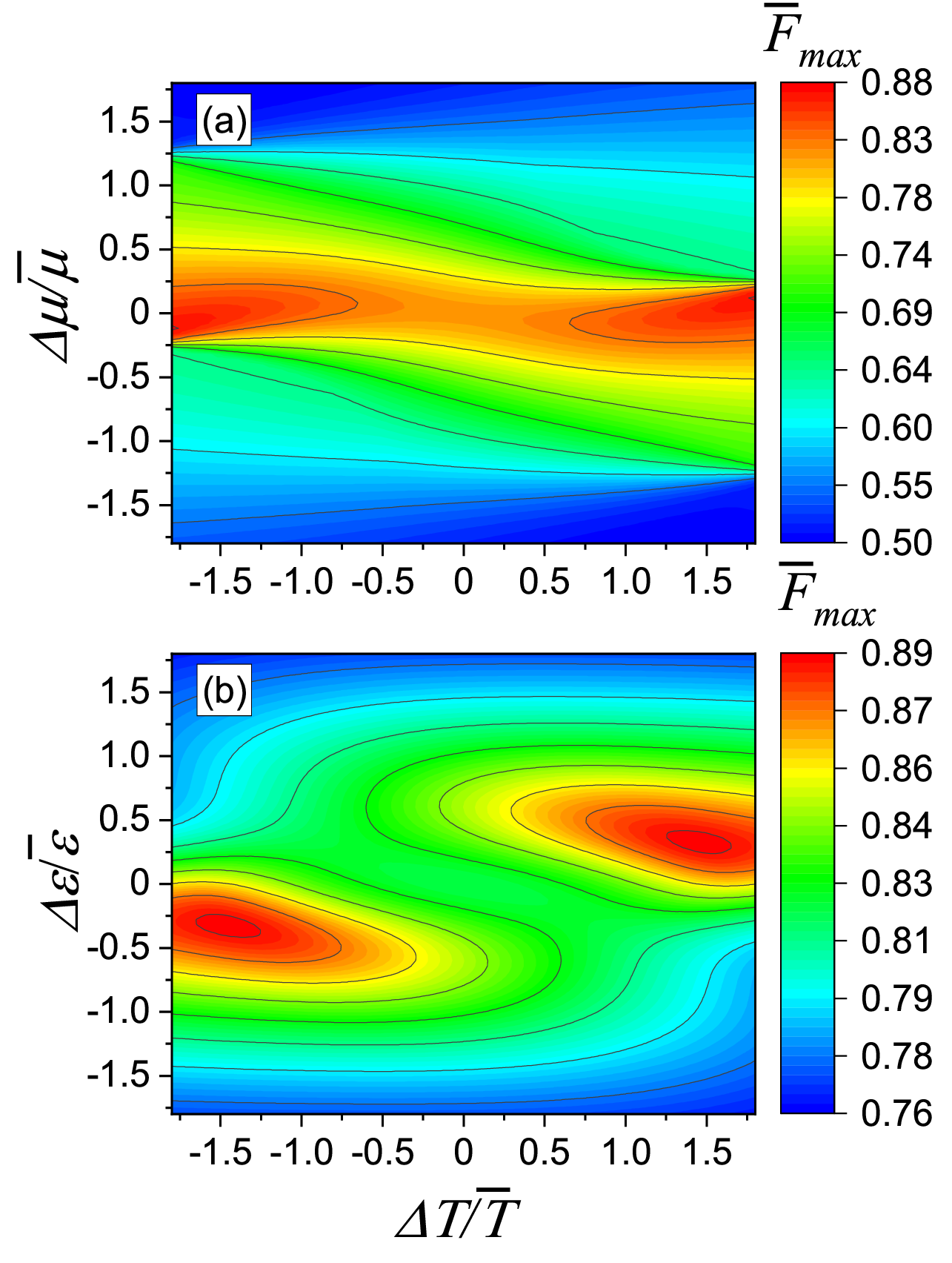}
    \caption{(a) The maximal average fidelity in terms of the non-equilibrium temperature difference \(\Delta T=T_A-T_B,\bar{T}=1\) and the chemical potential difference \(\Delta\mu=\mu_A-\mu_B,\bar{\mu}=8\). The energy levels are set as \(\varepsilon_A=\varepsilon_B=10\). (b) The maximal average fidelity in terms of the energy detuning \(\Delta\varepsilon=\varepsilon_A-\varepsilon_B\) with $\bar{\varepsilon}=10$ and the non-equilibrium temperature difference \(\Delta T=T_A-T_B\) with $\bar{T}=1$. The chemical potential is set as \(\mu_A=\mu_B=8\). Other parameters are set as \(\lambda=6 \) and \(g_A=g_B=0.05\).}
    \label{fig:ssftemu}
\end{figure}

From the results of the previous analysis, temperature difference \(\Delta T\) can enhance fidelity for selected chemical potential. When the temperature and chemical potential are both non-equilibrium, these non-equilibrium conditions have more subtle effects on fidelity. Fig. \ref{fig:ssftemu} illustrates the influence of  \( \Delta T / \bar{T} \) and \( \Delta \mu / \bar{\mu} \) on the \( \bar{F}_{max} \). Specific combinations of \( \Delta T \) and \( \Delta \mu \) result in significant enhancement of \( \bar{F}_{max} \), indicating the combined influence of temperature and chemical potential on the fidelity. High temperatures and chemical potentials can facilitate the excitation, while the other reservoir suppresses the excitation of the qubit. Consequently, the system is more likely to be in a single-excitation state. Since the single-excitation state includes entanglement, the fidelity is also enhanced accordingly. 

\begin{figure}[t!]
    \centering
    \includegraphics[width=0.5\textwidth]{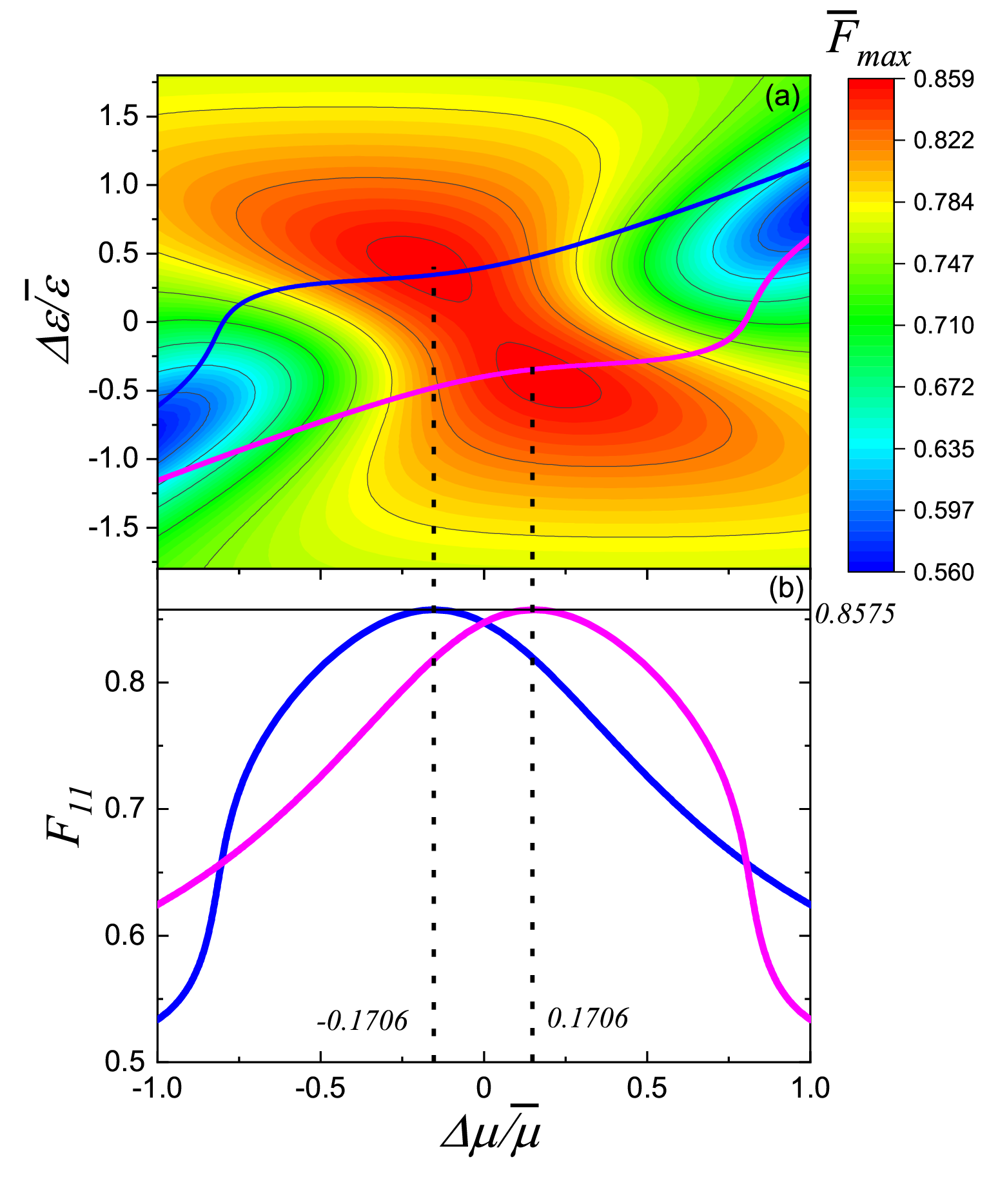}
    \caption{The color-filled contour map shows (a) the maximal fidelity in terms of the energy detuning \(\Delta \varepsilon=\varepsilon_A-\varepsilon_B\) and the non-equilibrium chemical potential difference \(\Delta \mu=\mu_A-\mu_B\). The two lines in (a) indicate the values of \(\Delta \varepsilon\) and \(\Delta \mu\) giving the fixed-point fidelity. The corresponding fidelity \(\bar{F}_{11}\) is shown in (b). The horizontal line at 0.8575 marks the peak of \(\bar{F}_{11}\), while the vertical dasheded line indicates the corresponding \(\Delta \mu\) value around \(\pm0.1706\). The parameter are set as \(T_A=T_B=1.5\), \(\varepsilon_A=\varepsilon_B=10\), \(\lambda=6\), and \(g_A=g_B=0.05\).}
    \label{fig:ssfmue1}
\end{figure}

In the fermionic reservoir scenario, the chemical potential difference and the asymmetric energy levels of the system can jointly enhance the fidelity, similar to the bosonic case, as shown in  Fig. \ref{fig:ssfmue1}. Although Fig. \ref{fig:ssfmue1} reveals a pattern similar to the bosonic reservoir case, the underlying mechanisms are fundamentally different. The detuning $|\Delta \varepsilon|$ changes the energy gap between the ground state and excited states. The lower energy level qubit coupled with the higher chemical potential reservoir can increase the population of the excited states and maintain entanglement. Fidelity is enhanced in these regions, particularly when $|\Delta \mu/\bar{\mu} |\approx 0.1706$, as shown in Fig. \ref{fig:ssfmue1} (a). Notably, fermionic reservoirs demonstrate an advantage: they achieve significant fidelity improvements at the relatively weak interaction phase, whereas bosonic reservoirs require much stronger interactions for comparable performance.





The fixed-point fidelity, previously found in the bosonic cases, also exists in the fermionic non-equilibrium setups. In Fig. \ref{fig:ssfmue1} (a), the blue and magenta lines indicate conditions at which $f_{11} = 0$, and $\rho_{AB}(1,4)=\rho_{AB}(4,1) = 0$. The latter is equivalent to \(\alpha=0\) in Eq. (\ref{rhoX}). Under these two conditions, the fidelity $F_{11}$ is independent of the input state parameters. The maximum $\bar{F}_{max}$ of those two lines is around 0.8575, while the maximum $\bar{F}_{max}$ of Fig. \ref{fig:ssfmue1} (a) is about 0.8583. Although the two fixed-point lines do not intersect the peak values in Fig. \ref{fig:ssfmue1} (a), they are fairly close. The mismatch between the maximal fixed-point fidelity and the overall fidelity is due to the non-zero imaginary component of the density matrix  $\rho_{AB}(2,3)$, where the $\cos\epsilon$ term in Eq. (\ref{F0111}) reduces the fidelity. The presence of this imaginary component reflects the quantum coherence induced by the chemical potential differences, which, while beneficial for entanglement generation, introduces a phase-dependent fidelity. These results demonstrate that careful tuning of chemical potential and system energy levels can simultaneously enhance fidelity and achieve a phase-dependent teleportation performance. 

In the strong coupling system within fermionic reservoirs, the ground state of the two-qubit system is entangled, providing a foundation for quantum teleportation. However, as the average temperature \(\bar{T}\) and average chemical potential \(\bar{\mu}\) increase, the system becomes susceptible to excitations, as illustrated in Fig. \ref{fig:l30_f}. Specifically, when the chemical potentials of the two baths are non-equilibrium, the fidelity remains suboptimal across both high and low temperature regimes, as evidenced by Fig. \ref{fig:l30_f} (a). Furthermore, in the presence of a temperature imbalance between the two baths, the overall fidelity tends to decrease with increasing chemical potential. Nevertheless, under certain conditions, an appropriate value of the chemical potential allows the temperature imbalance to enhance fidelity. This behavior is exemplified by the curves corresponding to \((\bar{T}=1,\mu=4)\), \((\bar{T}=5,\mu=1)\) and \((\bar{T}=5,\mu=3)\) in Fig. \ref{fig:l30_f} (b).

At low temperatures and low chemical potentials, the system predominantly resides in its ground state, with minimal population in excited states. However, a temperature difference, $\Delta T$, can still induce excitations: the cold reservoir facilitates relaxation to lower energy levels, while the hotter reservoir drives transitions to higher energy states. As the chemical potential increases, the population of the first excited state gradually rises, and the cooling effect of the cold reservoir becomes more significant. In this regime, $\Delta T$ effectively drives the system back toward the ground state, thereby enhancing entanglement and, consequently, improving the teleportation fidelity. However, once the chemical potential surpasses a critical threshold, the first excited state becomes densely populated, and the influence of the temperature difference diminishes. As a result, entanglement is significantly reduced, and the fidelity converges toward the classical limit of $2/3$.

When both the temperature and the chemical potential are driven out of equilibrium, the regions corresponding to maximum and minimum fidelity are situated in close proximity to one another, as depicted in Fig. \ref{fig:l30_fd}. This narrow separation highlights the extreme sensitivity of fidelity to simultaneous variations in these two parameters. In addition to the prominent red bands in the upper-left and lower-right corners, which signify significant fidelity enhancement, the lower-left and upper-right regions also exhibit slightly elevated fidelities compared to the central area. Within our model, once the chemical potential exceeds a threshold value of 5, the teleportation fidelity converges to the classical bound of $2⁄3$, and the quantum advantage diminishes to negligible levels. Consequently, the low–chemical-potential regime is not the primary focus of this study, particularly in the context of a fermionic thermal bath, as it offers limited insights into the quantum enhancement of teleportation. Moreover, there is no fixed-point fidelity phenomenon in this non-equilibrium environment.

\section{TELEPORTATION BY TRANSIENT STATE IN non-equilibrium ENVIRONMENTS }\label{s4}

Steady-state entanglement requires non-zero interaction strength between qubits, but it is not practical for two qubits that are located far apart. If the two qubits that do not interact are initially maximally entangled, the teleportation fidelity can still exceed the classical limit of $2/3$ in the presence of environment noises \cite{PhysRevLett.128.170501,xia2017long}. In this section, we focus on examining the fidelity of quantum teleportation by the transient state under the influence of non-equilibrium environments. We separately discuss the bosonic and fermionic environments in the following.

\subsection{Bosonic reservoirs}


Starting with two qubits initialized in state $|\beta(00)\rangle_{AB}$, environmental interactions lead to entanglement degradation and consequently the decreased teleportation fidelity. For our analysis with the initial state $|\beta(00)\rangle$, we employ the standard teleportation protocol (Bell measurements with Pauli corrections), in which the fidelity of state $|\psi\rangle_{\bar{A}}$ transmitted via the X-form entangled resource is given by Eq. (\ref{F0010}). The parameter $f_{00}$ defined in Eq. (\ref{f_00_10}) plays a crucial role in determining the fixed-point fidelity. When $f_{00}$ vanishes, fidelity becomes independent of the input state parameter \(\theta\).

\begin{figure}[t!]
    \centering
    \includegraphics[width=0.5\textwidth]{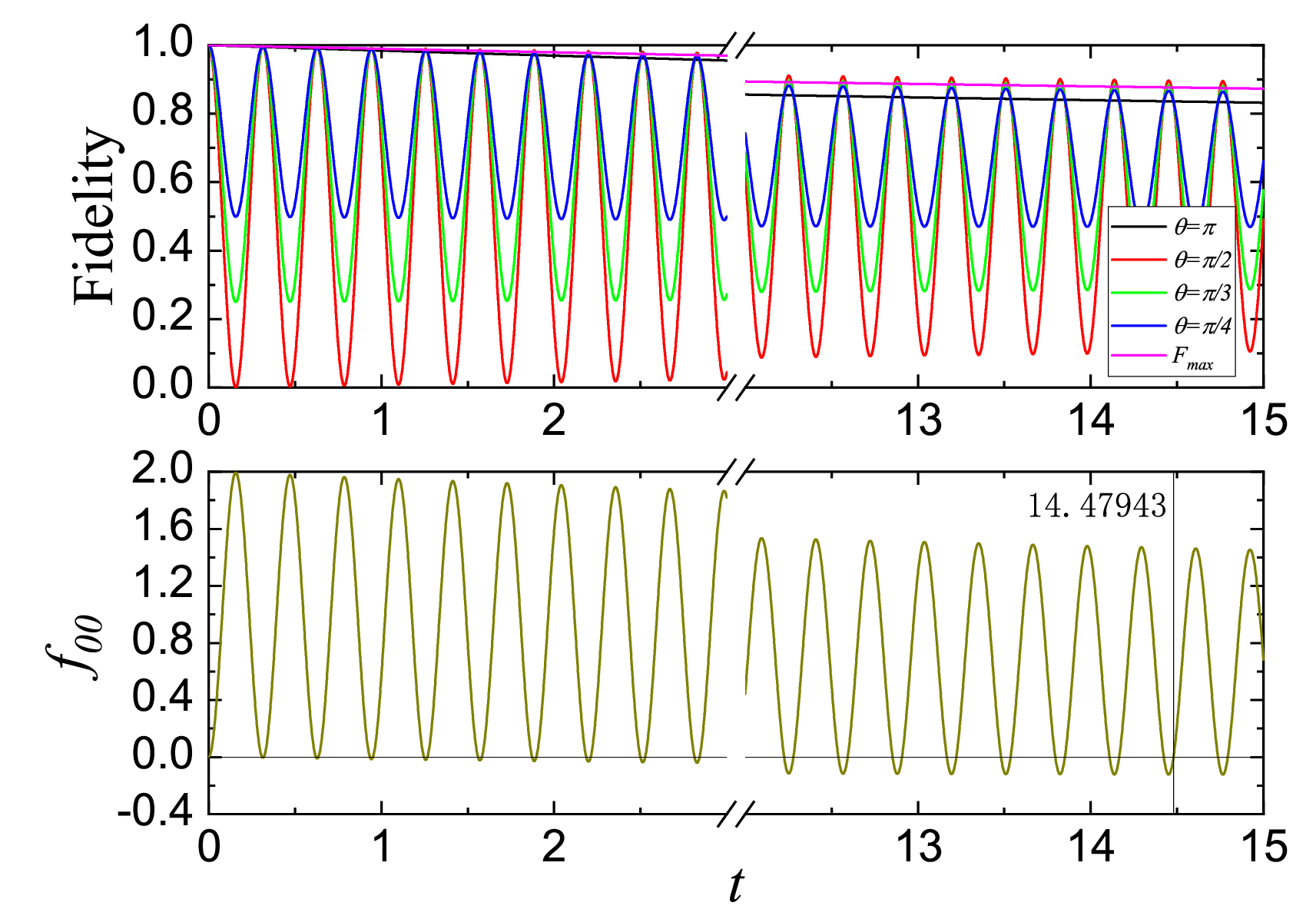}
    \caption{The fidelity \(F_{00}\) of different input states and the maximal fidelity \(\bar{F}_{max}\) in terms of the evolution time are shown in (a). The corresponding parameter \(f_{00}\) in \(F_{00}\) is shown in (b) and the vertical and horizontal coordinate lines indicate that \(f_{00}=0\) at \(t=14.479\). The initial entanglement resource is set as \(|\beta\rangle_{00}\). The qubits and the environments have the parameters $T_A = T_B = 2$, $\varepsilon_A = \varepsilon_B = 10$, $g_A = g_B = 0.05$, and $\lambda=0$.}
    \label{fig:tbeq}
\end{figure}

Under the Bloch-Redfield master equation, we calculate the transient state and its corresponding teleportation fidelities. Fig. \ref{fig:tbeq} illustrates the temporal decay of fidelity  under equilibrium conditions. Subfigure (a) shows fidelity oscillations over time for all $\theta$ values except $\theta=\pi$, with $\bar{F}_{max}$ exhibiting a gradual, monotonic decay. Subfigure (b) reveals $f_{00}$'s oscillatory behavior, periodically reaching zero. This behavior demonstrates the existence of fixed-point fidelity. Numerically we observe, $\alpha$ gradually decaying over time while $\beta$ increases, causing temporal oscillations in fidelity. Notably, in the non-interacting case $\lambda=0$, the parameter $\delta$ remains zero, making the fidelity $F_{00}$ independent of phase parameter $\phi$ of $|\psi\rangle_{\bar{A}}$. Consequently, when $f_{00}=0$, all arbitrary states input $|\psi\rangle_{\bar{A}}$ achieve identical fidelity values.

\begin{figure}[t!]
    \centering
    \includegraphics[width=0.5\textwidth]{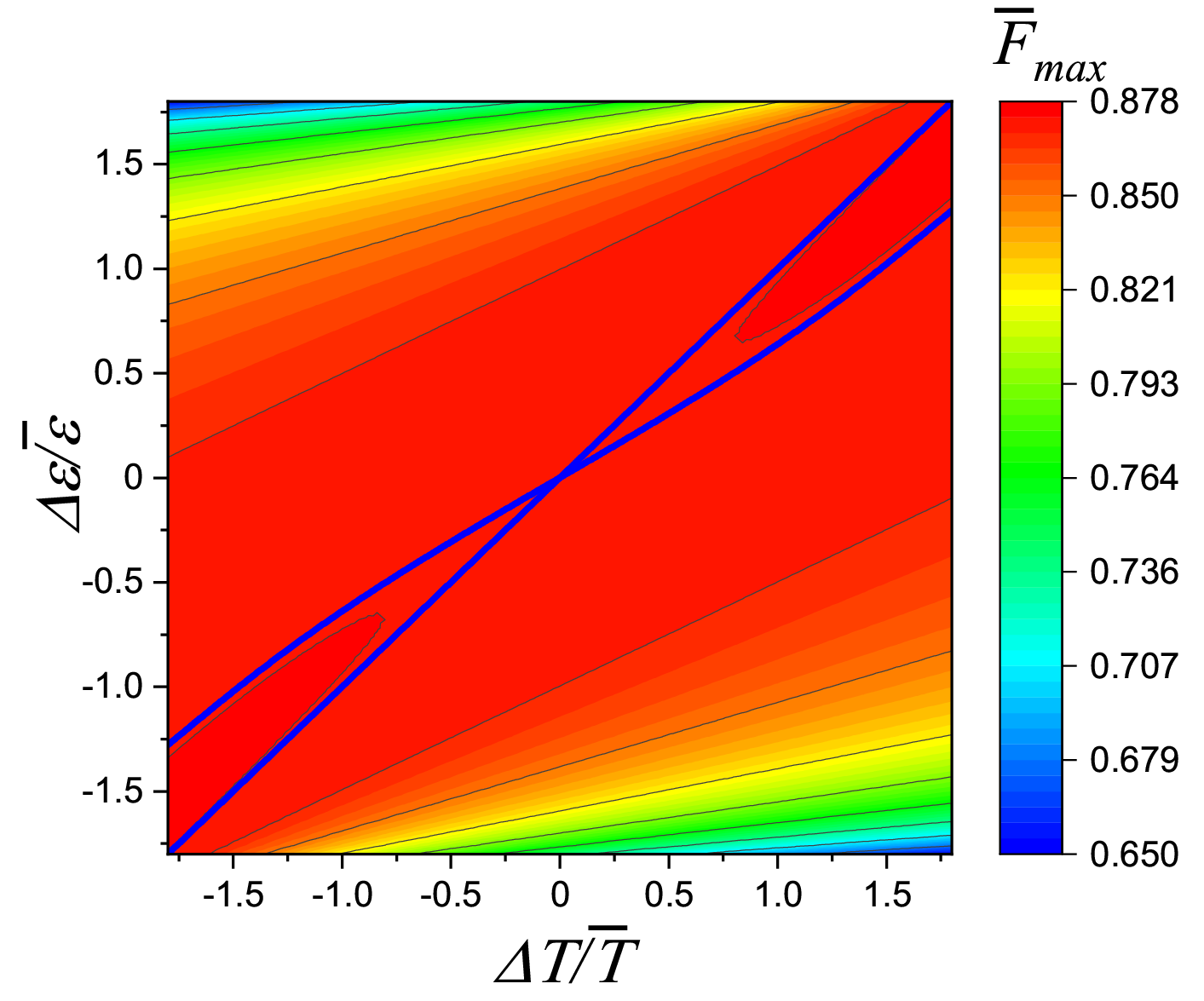}
    \caption{The color-filled contour map shows the maximal average fidelity in terms of the energy detuning \(\Delta \varepsilon=\varepsilon_A-\varepsilon_B\) and the non-equilibrium temperature difference \(\Delta T=T_A-T_B\) at \(t=14.479 \). The blue lines indicate the fixed-point fidelity under the condition \(f_{00}=0\). The parameters are set as $T_{A}=T_B=2$, $\varepsilon_{A}=\varepsilon_{B}=10$, $g_{A}=g_{B}=0.05$, and $\lambda=0$.}
    \label{fig:tbte}
\end{figure}

Under the combined non-equilibrium conditions (non-zero $\Delta \varepsilon$ and $\Delta T$), the fidelity shows behavior that closely resembles steady-state patterns. In Fig. \ref{fig:tbte}, we focus our analysis on the time point $t=14.479$, where $f_{00}$ reaches zero in Fig. \ref{fig:tbeq} (b).  The qubit with a higher energy level, when coupled to a higher-temperature reservoir, results in a modest enhancement of fidelity. The fixed-point fidelity, under the condition $f_{00} = 0$ (represented by the blue line), can also be enhanced due to the joint influences of the non-equilibrium reservoirs and the detuning of two qubits. We adopt the moment $t=14.479$ and employ the transient state in this moment to implement quantum teleportation. This setup can guarantee the fixed-point fidelity must appear in the equilibrium state.

The influence of non-equilibrium conditions on $f_{00}$ turns out to be relatively modest compared to its temporal oscillation amplitude. The influence of non-equilibrium conditions has important implications for fixed-point fidelity. If $f_{00}$ deviates significantly from zero, non-equilibrium conditions cannot compensate and restore it to the fixed-point fidelity. This behavior reveals a fundamental characteristic of fixed-point fidelity. In other words, fixed-point fidelity is highly localized, occurring only within narrow temporal windows. The behavior of fixed-point fidelity contrasts with that in the steady-state case, where appropriate non-equilibrium conditions can maintain fixed-point fidelity over extended periods.

\subsection{Fermionic reservoirs}

\begin{figure}[t!]
    \centering
    \includegraphics[width=0.5\textwidth]{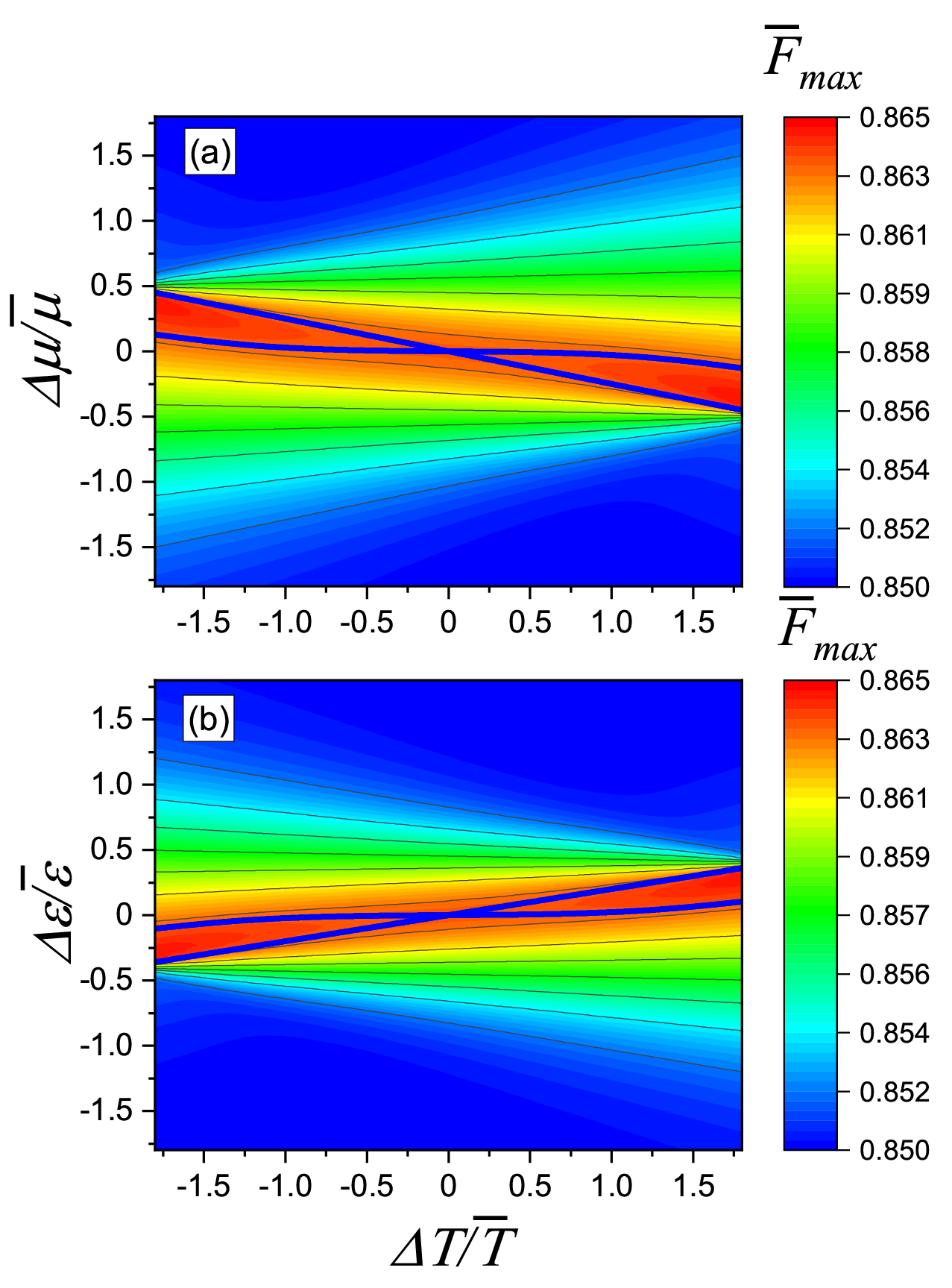}
    \caption{Maximal average fidelity in terms of the non-equilibrium temperature difference $\Delta T=T_A-T_B$ with $\bar{T}=1$ and (a) the non-equilibrium chemical potential difference $\Delta \mu=\mu_A-\mu_B$ with $\bar{\mu}=8$ or (b) the energy detuning $\Delta \varepsilon=\varepsilon_A-\varepsilon_B$ with $\bar{\varepsilon}=10$. The blue lines indicate the fixed-point fidelity condition $f_{00}=0$. The evolution time is $t=4.052$. The other parameters are set as $g_{A}=g_{B}=0.1$ and $\lambda=0$.}
    \label{fig:tftmu}
\end{figure}

In this section, we examine the fidelity when the system is coupled to the fermionic reservoirs. Similar to the bosonic case, we set the initial state as $|\beta(00)\rangle_{AB}$. The fidelity also exhibits temporal oscillations, similar to Fig. \ref{fig:tbeq}. Next, we analyze the specific time point \(t=4.052\) when $f_{00}$ vanishes in the equilibrium case, which corresponds to a fixed-point fidelity. Fig. \ref{fig:tftmu} (a) reveals the fidelity as a function of the non-equilibrium temperature and chemical potential differences. The fidelity $\bar{F}_{max}$ is enhanced if the high-temperature reservoir is coupled with the lower chemical potential reservoir, and vice versa. These enhancements reach their peak under extreme non-equilibrium conditions of $\Delta T$ and $\Delta \mu$. 

Fig. \ref{fig:tftmu} (b) shows the influence of the detuning energy and the temperature difference on fidelity. The qubit with a higher energy level, when coupled to a higher-temperature reservoir, results in a modest enhancement of fidelity, and vice versa. Blue lines across both figures illustrate the specific configurations satisfying the fixed-point condition $f_{00} = 0$. The fixed-point fidelity can also be enhanced by the non-equilibrium conditions, similar to the maximal average fidelity. 

The transient states shown in Fig. \ref{fig:tftmu} (a) and (b) have higher fidelity \(F_{max}\) in the region with lower concurrence. At this point, we cannot simply explain the distribution of fidelity in the graph by the strength of concurrence; rather, we need to further discuss the relationship between fidelity and the density matrix. More detailed data and analysis will be provided in the Appendix \ref{A}.


\begin{figure}[t!]
    \centering
    \includegraphics[width=0.5\textwidth]{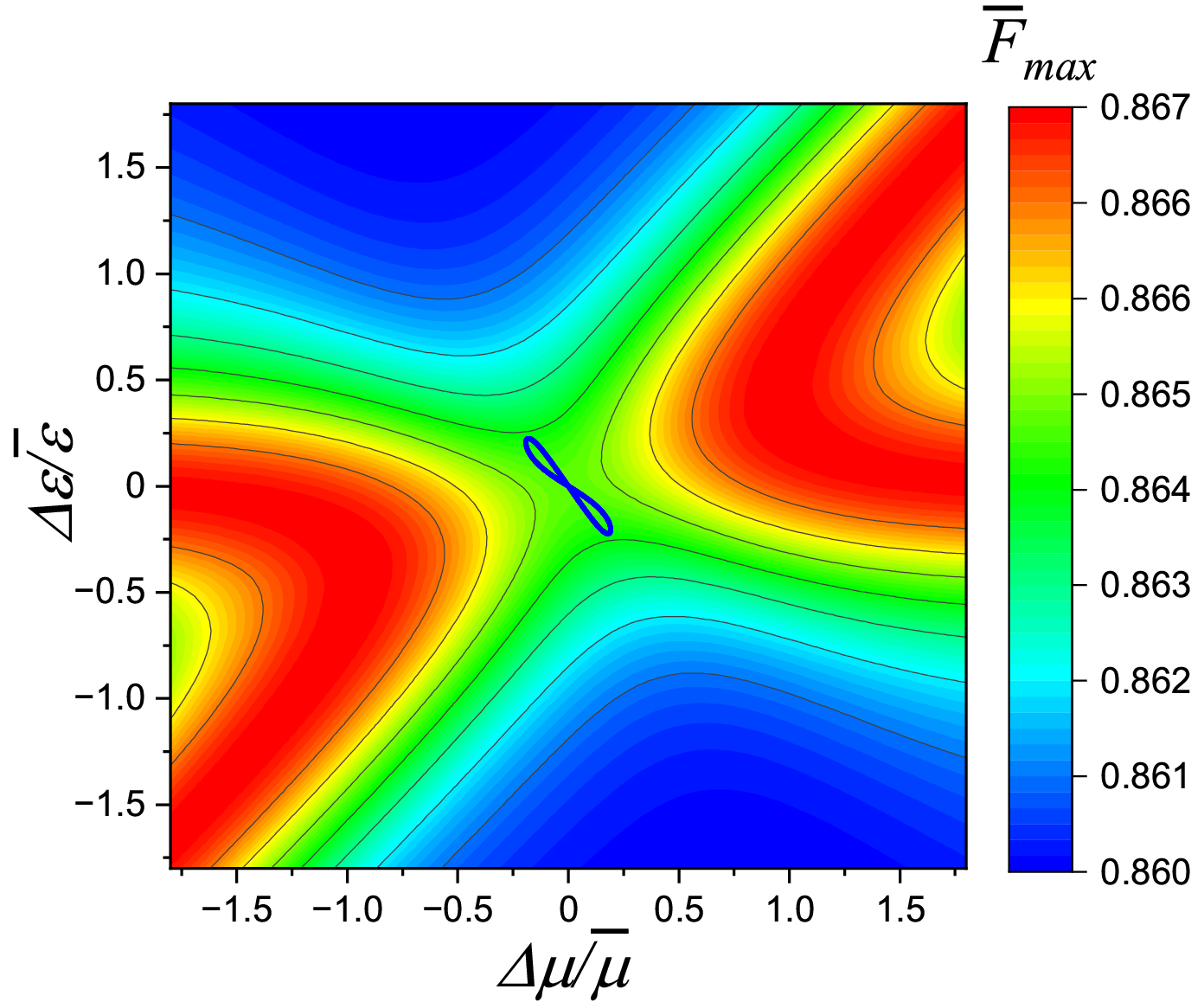}
    \caption{The color-filled contour map shows the maximal fidelity in terms of the energy detuning \(\Delta \varepsilon=\varepsilon_A-\varepsilon_B\) and the non-equilibrium chemical potential difference \(\Delta \mu=\mu_A-\mu_B\). The coupled time is set as \(t=3.751 \). The blue line indicates the fixed-point fidelity condition \(f_{00}=0 \). Other parameters are set as $T_{A}=T_B=2$, $\mu_A=\mu_B=10$, $\varepsilon_{A}=\varepsilon_{B}=10$, $g_{A}=g_{B}=0.1$, and $\lambda=0$.}
    \label{fig:tftemuC10}
\end{figure}

As for the fidelity in terms of the detuning $\Delta \varepsilon$ and the non-equilibrium condition $\Delta\mu$, as shown by Fig. \ref{fig:tftemuC10}. The maximal average fidelity $\bar{F}_{max}$ now exhibits a more complex distribution characterized by two distinct band-like regions with enhanced fidelity. These enhancement zones emerge under the conditions that the qubit with a higher (lower) energy level should couple to the reservoir with a higher (lower) chemical potential. The interactions with non-equilibrium environments introduce rich complexity into the fidelity landscape. From Appendix \ref{A}, $\bar{F}_{max}$ is determined by concurrence. The blue line in Fig. \ref{fig:tftemuC10} shows the fixed-point fidelity condition \(f_{00}=0\). In contrast to the linear behavior in Fig. \ref{fig:tftmu}, it has a twisted ring localized in the vicinity of the origin. However, the ring configuration does not overlap with the area where the maximal fidelity is enhanced. In other words, if we select the energy levels and non-equilibrium chemical potentials giving the enhanced fidelity $\bar{F}_{max}$, the specific fidelity inevitably depends on the input parameter \(\phi\) because the density matrix term \(\rho_{23}\) in Eq. (\ref{rhoX}) does not vanish.

\section{Conclusion}\label{s6}

We investigate quantum teleportation by a two-qubit system coupled with the non-equilibrium bosonic or fermionic reservoirs. We apply the Bloch-Redfield equation to describe the non-equilibrium dynamics and consider both the steady state and the transient state as the entanglement resource. The fidelity can be enhanced if the qubit with high-energy level couples to the high-temperature reservoir assuming the two-qubit system is in the strong coupling phase. The optimal strategy is dependent on the interaction strength \(\lambda\), which determines the energy level structure and the ground state entanglement properties. We also extend our steady-state analysis to the transient state with the Bell state \(|\beta(00)\rangle\) as the initial state. Our results reveal that the fidelity enhancement under similar parametric conditions exhibits patterns analogous to those observed in the steady-state regime. An interesting phenomenon is the teleportation exhibiting fixed-point fidelity. It means the fidelity is independent of the parameters of the input states. We find that the fixed-point fidelity can exist in steady state and transient state. Moreover, by combining the non-equilibrium condition with the detuning control, the fixed-point fidelity can be enhanced beyond the equilibrium values.

Our results enrich the understanding of quantum teleportation in realistic environments and provide practical guidance for implementing quantum communication protocols in the presence of reservoirs \cite{PhysRevA.67.032302,pirandola2021environment,sun2016quantum}. The fixed-point fidelity mechanism offers a promising pathway toward protocol simplification in practical quantum teleportation schemes, representing a potential application in quantum information science. 

\quad

\begin{acknowledgments}
X. K. Yan thanks NSF 12234019 for support. K. Zhang is supported by the National Natural Science Foundation of China under Grant Nos. 12305028 and 12247103, China Postdoctoral Science Foundation under Grant Number 2025M773421, Scientific Research Program Funded by Education Department of Shaanxi Provincial Government (Program No.24JP186), and the Youth Innovation Team of Shaanxi Universities.
\end{acknowledgments}

\appendix

\section{The entanglement of non-equilibrium 
quantum states}\label{A}

In this appendix, we aim to illustrate how entanglement performance under the non-equilibrium conditions discussed in the main text, which offers a more intuitive understanding of the relationship between fidelity and entanglement. While detailed analysis, including both steady state and dynamic cases, has been extensively explored in previous studies, we will focus here solely on demonstrating selected scenarios as outlined in this work. Here we apply the concurrence as the entanglement measure \cite{PhysRevLett.78.5022}, and for a two-qubit system concurrence is given by 
\begin{align}
    C(\rho)=\text{Max}(0,\sqrt{\zeta_1}-\sqrt{\zeta_2}-\sqrt{\zeta_3}-\sqrt{\zeta_4}),
\end{align}
where \(\zeta_i\) are the eigenvalues of the matrix \(\rho(\sigma_y\otimes\sigma_y)\rho^*(\sigma_y\otimes\sigma_y)\) in descending order. For the X-state in our study, it can be easily calculated that the concurrence reduces to the following simple expression:
\begin{align}
    C(\rho^X)=2\text{Max}(0, \delta-\sqrt{ad},\alpha-\sqrt{bc}).
\end{align}
Compared with Eq. (\ref{FmaxX}), one can conclude that the concurrence cannot determine the fidelity completely, but in most cases, the fidelity and concurrence exhibit a positive correlation \cite{nandi2018two}. In this paper, we have observed a positive correlation, or minor violations thereof, between fidelity and concurrence in teleportation protocols during the steady state and transient state bosonic reservoirs. To validate the accuracy of our fidelity analysis presented in the main text, we will now demonstrate the variation of concurrence in the steady state process when a single parameter is out of equilibrium. Furthermore, we will also discuss the concurrence behavior during the evolution process.

\begin{figure}[htbp]
    \centering
    \includegraphics[width=0.5\textwidth]{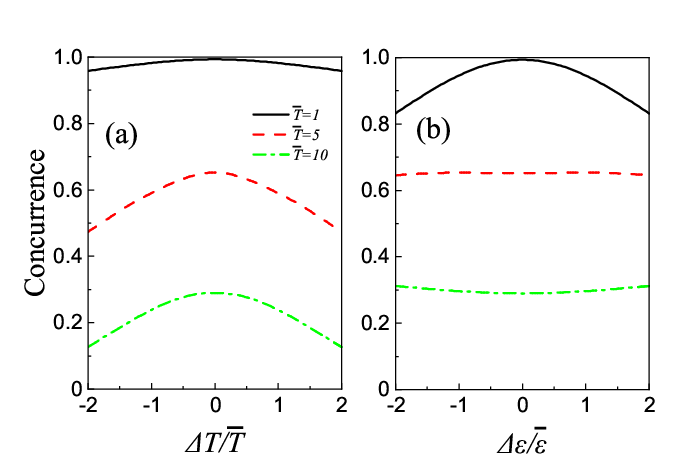}
    \caption{Concurrence of the steady state couple to the bosonic environments with non-equilibrium condition \(\Delta T= T_A - T_B\) for (a) and energy level asymmetric \(\Delta \varepsilon= \varepsilon_A - \varepsilon_B\). The average temperature set as low temperature \(\bar{T}=1\) (black solid line), moderate temperature \(\bar{T}=5\) (red dashed line) and high temperature \(\bar{T}=10\) (green dashed dot line). The average energy is \(\bar{\varepsilon}=10\). Other parameter are set as \(\lambda=30\) and \(g_A=g_B=0.05\).}
    \label{fig:dT}
\end{figure}

For the case of bosonic reservoirs, high temperature drives the system into excited states and affects entanglement in different ways when interaction strength across different ranges. In the setup of Fig. \ref{fig:bssT}, we illustrate the variation of fidelity with temperature difference and detuning energy levels for \( \lambda = 30 \).  When \( \lambda = 30 \), the ground state \(|2\rangle\) is entangled, and both an increase in mean temperature and the temperature difference enhance the excitation of the system, consequently leading to the degradation of entanglement as shown in Fig. \ref{fig:dT} (a).

The fidelity evolution under energy detuning \(\Delta\varepsilon\) is presented in Fig. \ref{fig:bssT} (b), while the corresponding concurrence dynamics are shown in Fig. \ref{fig:dT} (b). The impact of \(\Delta\varepsilon\) varies with temperature. At low temperatures, the effect of \(\Delta\varepsilon\) primarily manifests as a reduction in ground state entanglement shown by the black solid line in Fig. \ref{fig:dT} (b). The asymmetric energy levels \(\Delta\varepsilon\) not only suppress system excitation but also reduce the ground state entanglement; however, the overall effect is a decrease in entanglement. When \(\bar{T}=5,10\) (red dashed line and green dash-dot line in Fig. \ref{fig:dT}(b)), thermal excitation dominates the system dynamics. Here, \(\Delta\varepsilon\) primarily contributes to the resistance against thermal excitation.

\begin{figure}[h]
    \centering
    \includegraphics[width=0.5\textwidth]{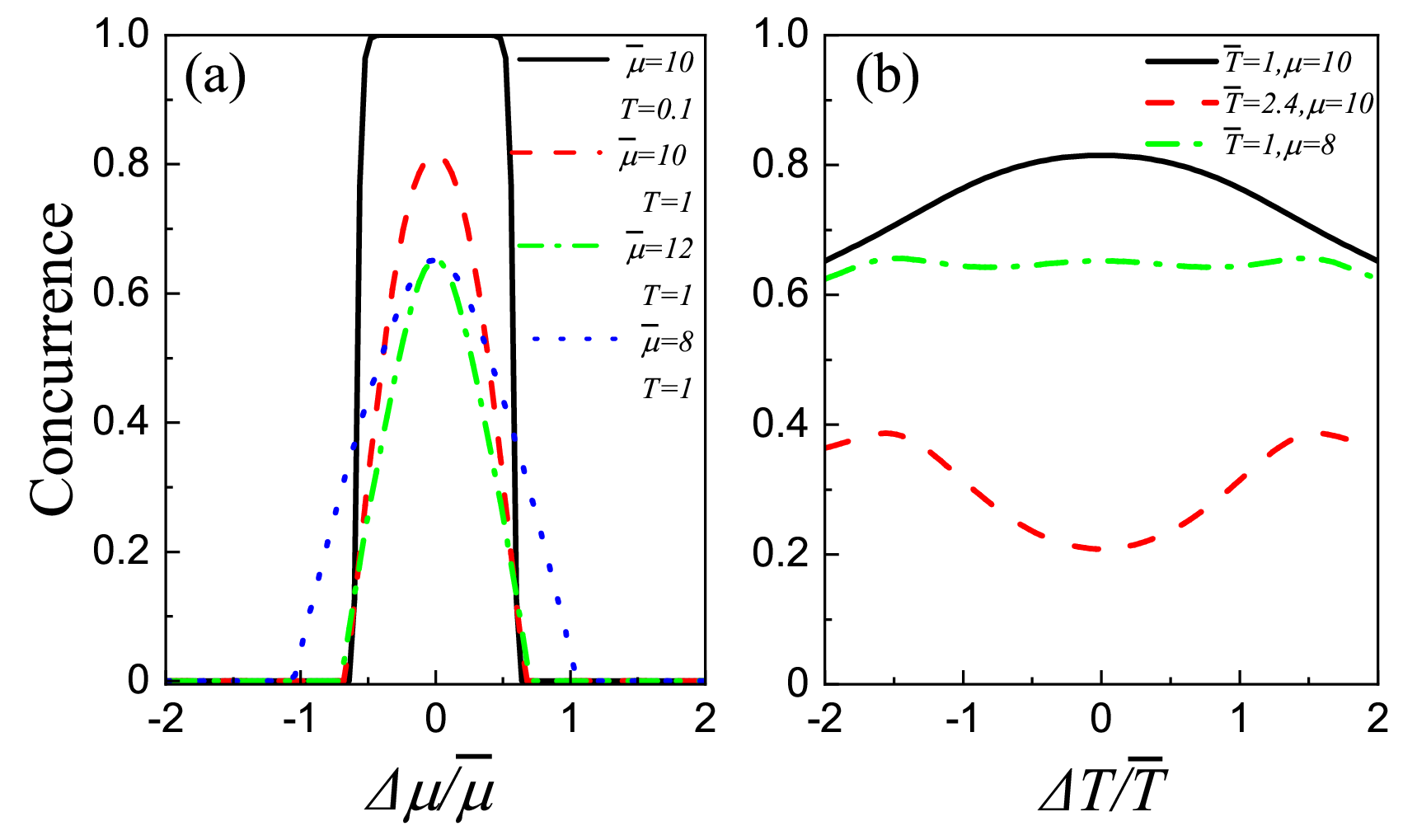}
    \caption{Concurrence of the system couple with fermionic environments with non-equilibrium chemical potential  \(\Delta \mu=\mu_A-\mu_B\) for (a) and temperature \(\Delta T=T_A-T_B\) for (b). In (a), the parameters set as \(\bar{\mu}=10,T=0.1\) (black solid line), \(\bar{\mu}=10,T=1\) (red dashed line), \(\bar{\mu}=12,T=1\) (green dash-dot line) and \(\bar{\mu}=8,T=1\) (blue dot line). In (b), the parameters set as \(\bar{T}=1,\mu=10\) (black solid line), \(\bar{T}=2.4,\mu=10\) (red dashed line) and \(\bar{T}=1,\mu=8\) (green dash-dot line). Other parameters are set as \(\varepsilon_A=\varepsilon_B=10\), \(\lambda=6 \) and \(g_A=g_B=0.05\).}
    \label{fig:dmu}
\end{figure}

In the case of fermionic reservoirs, the effects of energy detuning on entanglement exhibit similar patterns to those observed with bosonic reservoirs. In our analysis, which focuses on the weak interaction regime, the ground state \(|1\rangle\) is a product state. The chemical potential dominates the excitation of the system and leads to maximum concurrence, shown by the black solid line in Fig. \ref{fig:dmu} (a). Both the increase in average temperature and the deviation of average chemical potential from the resonant point can decrease the entanglement corresponding to the other lines in Fig. \ref{fig:dmu} (a).

The effect of temperature difference is shown in Fig. \ref{fig:dmu} (b). In the resonant point, shown as the black solid line and green dash-dot line in Fig. \ref{fig:dmu} (b), the temperature difference always reduces the concurrence around the equilibrium condition by destroying the maximum configuration of the population of state \(|2\rangle\) and \(|3\rangle\). When \(\bar{\mu}=8\), the temperature difference can offset the absence of excitation when deviating from the resonant point and enhance the concurrence shown by the green dash-dot line in Fig. \ref{fig:dmu} (b).

\begin{figure}[htbp]
    \centering
    \includegraphics[width=0.35\textwidth]{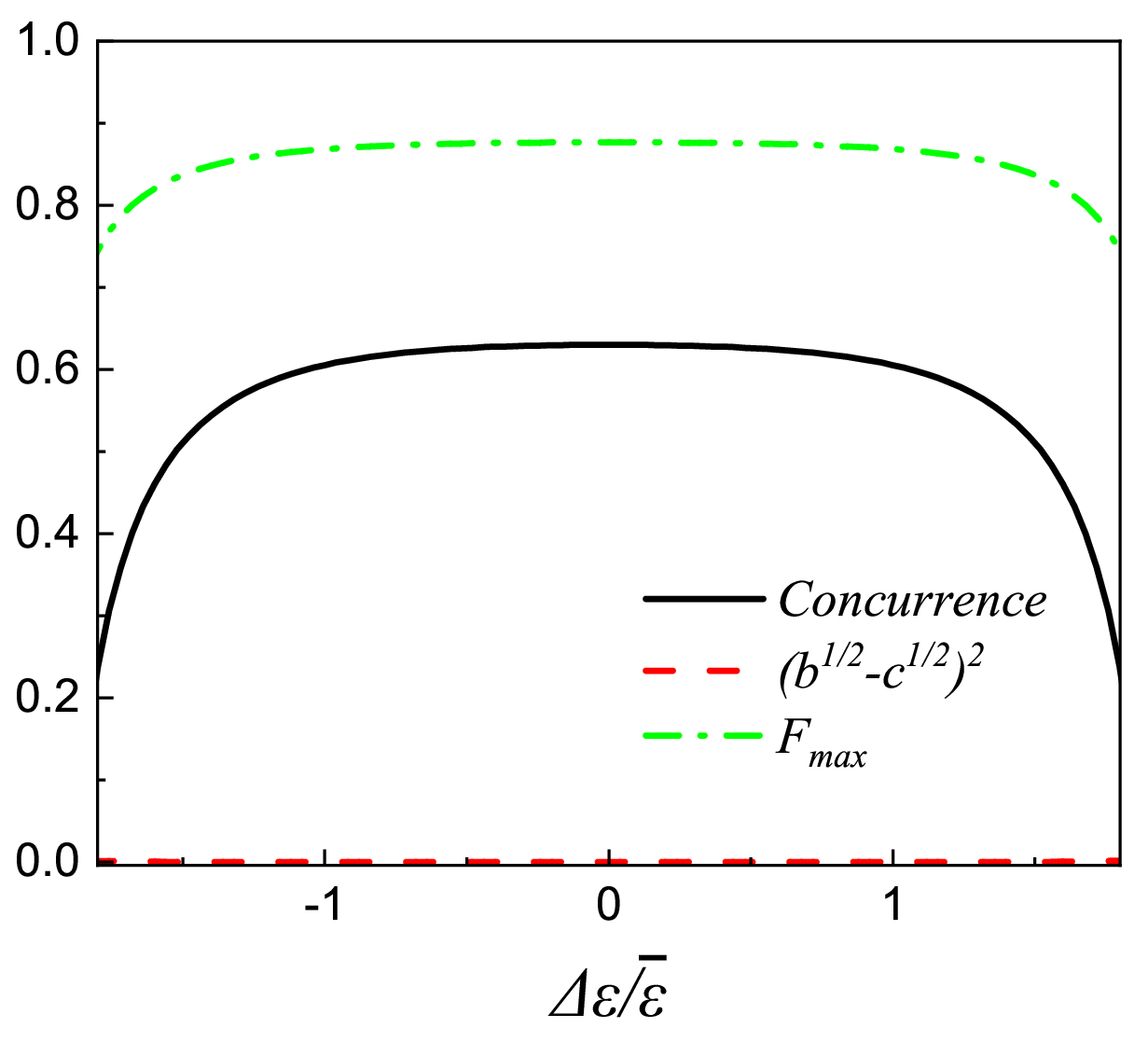}
    \caption{Concurrence and \(F_{max}\) of the transient state in bosonic reservoirs. This figure shows the concurrence (black solid lin), \((\sqrt{b}-\sqrt{c})^2\) (red dashed line) and \(F_{max}\) (green dash-dot line) corresponding to Fig. \ref{fig:tbte} with \(\Delta T/\bar{T}=0\).}
    \label{fig:btede}
\end{figure}

The above discussion is about how individual non-equilibrium conditions affect entanglement. When these conditions occur simultaneously, the behavior of entanglement becomes more complex due to the interplay of reinforcement and opposition among these influences \cite{zhang2021entanglement,wang2018steady}.

Considering the evolution process, the relationship between concurrence and fidelity becomes more complex, as the elements of the density matrix will oscillate over time under the Redfield equation. At the moments of interest—specifically, when fixed points occur—the density matrix elements remain relatively stable across different initial conditions. At this point, the relationship between concurrence and fidelity can be established. Because the initial state is set as \(|\beta (00)\rangle\), the density matrix element \(\rho(2,3)=\rho(3,2)=0\) imples that the concurrence is in the form as \(\alpha-\sqrt{bc}\). In our case the fidelity can be expressed as 
\begin{align}
     F_{max}=\frac{1}{3}(2+C-(\sqrt{b}-\sqrt{c})^2),
\end{align}
which is consistent with the numerical results after verification.

\begin{figure}[htbp]
    \centering
    \includegraphics[width=0.5\textwidth]{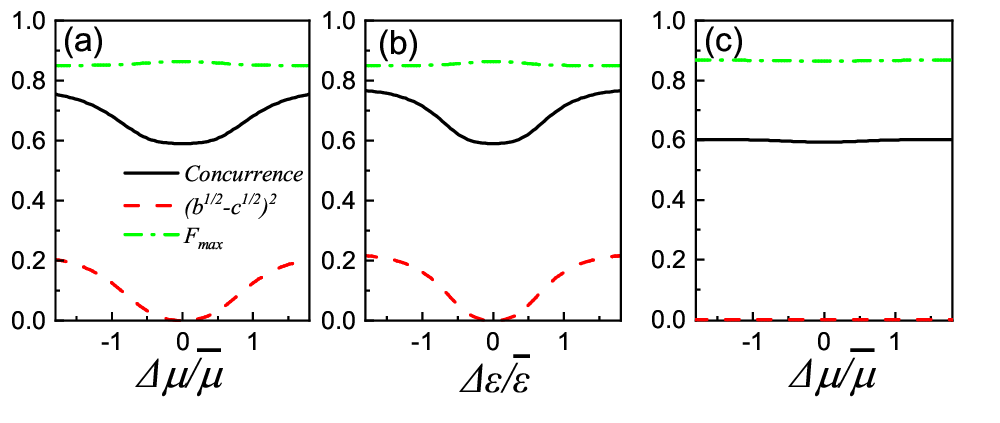}
    \caption{Concurrence and \(F_{max}\) of the transient state in fermionic reservoirs. This figure shows the concurrence (black solid lin), \((\sqrt{b}-\sqrt{c})^2\) (red dashed line) and \(F_{max}\) (green dash-dot line) corresponding to Fig. \ref{fig:tbte} with \(\Delta T/\bar{T}=0\). The subfigures (a) and (b) in Fig. \ref{fig:fcf} correspond to that in Fig. \ref{fig:tftmu} with fixed \(\Delta T/\bar{T}=0\) and the subfigure (c) corresponds to Fig. \ref{fig:tftemuC10} with fixed \(\Delta \varepsilon/\bar{\varepsilon}=0\).}
    \label{fig:fcf}
\end{figure}

To simultaneously compare the effects of concurrence and \((\sqrt{b}-\sqrt{c})^2\) on \(F_{max}\), we use a curve plot. In the case of bosonic reservoirs, as shown in Fig. \ref{fig:btede}, with fixed \(\Delta T/\bar{T}=0\), corresponding to Fig. \ref{fig:tbte}, the relatively small value of \((\sqrt{b}-\sqrt{c})^2\) compared to concurrence does not significantly impact the \(F_{max}\), and \(F_{max}\) has a positive correlation with concurrence.  

However, for the case of fermionic reservoirs, we consider three different configurations of non-equilibrium conditions and the subfigures (a) and (b) in Fig. \ref{fig:fcf} correspond to those in Fig. \ref{fig:tftmu} with fixed \(\Delta T/\bar{T}=0\) and the subfigure (c) corresponds to Fig. \ref{fig:tftemuC10} with fixed \(\Delta \varepsilon/\bar{\varepsilon}=0\). \((\sqrt{b}-\sqrt{c})^2\) becomes large enough and leads to an opposite distribution pattern for \(F_{max}\) compared to concurrence, as shown in Fig. \ref{fig:fcf} (a) and (b). When \((\sqrt{b}-\sqrt{c})^2\) is small, \(F_{max}\) has a positive correlation with concurrence.


\begin{thebibliography}{73}%
\makeatletter
\providecommand \@ifxundefined [1]{%
 \@ifx{#1\undefined}
}%
\providecommand \@ifnum [1]{%
 \ifnum #1\expandafter \@firstoftwo
 \else \expandafter \@secondoftwo
 \fi
}%
\providecommand \@ifx [1]{%
 \ifx #1\expandafter \@firstoftwo
 \else \expandafter \@secondoftwo
 \fi
}%
\providecommand \natexlab [1]{#1}%
\providecommand \enquote  [1]{``#1''}%
\providecommand \bibnamefont  [1]{#1}%
\providecommand \bibfnamefont [1]{#1}%
\providecommand \citenamefont [1]{#1}%
\providecommand \href@noop [0]{\@secondoftwo}%
\providecommand \href [0]{\begingroup \@sanitize@url \@href}%
\providecommand \@href[1]{\@@startlink{#1}\@@href}%
\providecommand \@@href[1]{\endgroup#1\@@endlink}%
\providecommand \@sanitize@url [0]{\catcode `\\12\catcode `\$12\catcode `\&12\catcode `\#12\catcode `\^12\catcode `\_12\catcode `\%12\relax}%
\providecommand \@@startlink[1]{}%
\providecommand \@@endlink[0]{}%
\providecommand \url  [0]{\begingroup\@sanitize@url \@url }%
\providecommand \@url [1]{\endgroup\@href {#1}{\urlprefix }}%
\providecommand \urlprefix  [0]{URL }%
\providecommand \Eprint [0]{\href }%
\providecommand \doibase [0]{https://doi.org/}%
\providecommand \selectlanguage [0]{\@gobble}%
\providecommand \bibinfo  [0]{\@secondoftwo}%
\providecommand \bibfield  [0]{\@secondoftwo}%
\providecommand \translation [1]{[#1]}%
\providecommand \BibitemOpen [0]{}%
\providecommand \bibitemStop [0]{}%
\providecommand \bibitemNoStop [0]{.\EOS\space}%
\providecommand \EOS [0]{\spacefactor3000\relax}%
\providecommand \BibitemShut  [1]{\csname bibitem#1\endcsname}%
\let\auto@bib@innerbib\@empty
\bibitem [{\citenamefont {Bennett}\ \emph {et~al.}(1993)\citenamefont {Bennett}, \citenamefont {Brassard}, \citenamefont {Cr\'epeau}, \citenamefont {Jozsa}, \citenamefont {Peres},\ and\ \citenamefont {Wootters}}]{PhysRevLett.70.1895}%
  \BibitemOpen
  \bibfield  {author} {\bibinfo {author} {\bibfnamefont {C.~H.}\ \bibnamefont {Bennett}}, \bibinfo {author} {\bibfnamefont {G.}~\bibnamefont {Brassard}}, \bibinfo {author} {\bibfnamefont {C.}~\bibnamefont {Cr\'epeau}}, \bibinfo {author} {\bibfnamefont {R.}~\bibnamefont {Jozsa}}, \bibinfo {author} {\bibfnamefont {A.}~\bibnamefont {Peres}},\ and\ \bibinfo {author} {\bibfnamefont {W.~K.}\ \bibnamefont {Wootters}},\ }\bibfield  {title} {\bibinfo {title} {Teleporting an unknown quantum state via dual classical and einstein-podolsky-rosen channels},\ }\href@noop {} {\bibfield  {journal} {\bibinfo  {journal} {Phys. Rev. Lett.}\ }\textbf {\bibinfo {volume} {70}},\ \bibinfo {pages} {1895} (\bibinfo {year} {1993})}\BibitemShut {NoStop}%
\bibitem [{\citenamefont {Pirandola}\ \emph {et~al.}(2015)\citenamefont {Pirandola}, \citenamefont {Eisert}, \citenamefont {Weedbrook}, \citenamefont {Furusawa},\ and\ \citenamefont {Braunstein}}]{pirandola2015advances}%
  \BibitemOpen
  \bibfield  {author} {\bibinfo {author} {\bibfnamefont {S.}~\bibnamefont {Pirandola}}, \bibinfo {author} {\bibfnamefont {J.}~\bibnamefont {Eisert}}, \bibinfo {author} {\bibfnamefont {C.}~\bibnamefont {Weedbrook}}, \bibinfo {author} {\bibfnamefont {A.}~\bibnamefont {Furusawa}},\ and\ \bibinfo {author} {\bibfnamefont {S.~L.}\ \bibnamefont {Braunstein}},\ }\bibfield  {title} {\bibinfo {title} {Advances in quantum teleportation},\ }\href@noop {} {\bibfield  {journal} {\bibinfo  {journal} {Nature photonics}\ }\textbf {\bibinfo {volume} {9}},\ \bibinfo {pages} {641} (\bibinfo {year} {2015})}\BibitemShut {NoStop}%
\bibitem [{\citenamefont {Bowen}\ and\ \citenamefont {Bose}(2001)}]{bowen2001teleportation}%
  \BibitemOpen
  \bibfield  {author} {\bibinfo {author} {\bibfnamefont {G.}~\bibnamefont {Bowen}}\ and\ \bibinfo {author} {\bibfnamefont {S.}~\bibnamefont {Bose}},\ }\bibfield  {title} {\bibinfo {title} {Teleportation as a depolarizing quantum channel, relative entropy, and classical capacity},\ }\href@noop {} {\bibfield  {journal} {\bibinfo  {journal} {Physical Review Letters}\ }\textbf {\bibinfo {volume} {87}},\ \bibinfo {pages} {267901} (\bibinfo {year} {2001})}\BibitemShut {NoStop}%
\bibitem [{\citenamefont {Verstraete}\ and\ \citenamefont {Verschelde}(2003)}]{verstraete2003optimal}%
  \BibitemOpen
  \bibfield  {author} {\bibinfo {author} {\bibfnamefont {F.}~\bibnamefont {Verstraete}}\ and\ \bibinfo {author} {\bibfnamefont {H.}~\bibnamefont {Verschelde}},\ }\bibfield  {title} {\bibinfo {title} {Optimal teleportation with a mixed state of two qubits},\ }\href@noop {} {\bibfield  {journal} {\bibinfo  {journal} {Physical review letters}\ }\textbf {\bibinfo {volume} {90}},\ \bibinfo {pages} {097901} (\bibinfo {year} {2003})}\BibitemShut {NoStop}%
\bibitem [{\citenamefont {Bouwmeester}\ \emph {et~al.}(1997)\citenamefont {Bouwmeester}, \citenamefont {Pan}, \citenamefont {Mattle}, \citenamefont {Eibl}, \citenamefont {Weinfurter},\ and\ \citenamefont {Zeilinger}}]{bouwmeester1997experimental}%
  \BibitemOpen
  \bibfield  {author} {\bibinfo {author} {\bibfnamefont {D.}~\bibnamefont {Bouwmeester}}, \bibinfo {author} {\bibfnamefont {J.-W.}\ \bibnamefont {Pan}}, \bibinfo {author} {\bibfnamefont {K.}~\bibnamefont {Mattle}}, \bibinfo {author} {\bibfnamefont {M.}~\bibnamefont {Eibl}}, \bibinfo {author} {\bibfnamefont {H.}~\bibnamefont {Weinfurter}},\ and\ \bibinfo {author} {\bibfnamefont {A.}~\bibnamefont {Zeilinger}},\ }\bibfield  {title} {\bibinfo {title} {Experimental quantum teleportation},\ }\href@noop {} {\bibfield  {journal} {\bibinfo  {journal} {Nature}\ }\textbf {\bibinfo {volume} {390}},\ \bibinfo {pages} {575} (\bibinfo {year} {1997})}\BibitemShut {NoStop}%
\bibitem [{\citenamefont {Boschi}\ \emph {et~al.}(1998)\citenamefont {Boschi}, \citenamefont {Branca}, \citenamefont {De~Martini}, \citenamefont {Hardy},\ and\ \citenamefont {Popescu}}]{PhysRevLett.80.1121}%
  \BibitemOpen
  \bibfield  {author} {\bibinfo {author} {\bibfnamefont {D.}~\bibnamefont {Boschi}}, \bibinfo {author} {\bibfnamefont {S.}~\bibnamefont {Branca}}, \bibinfo {author} {\bibfnamefont {F.}~\bibnamefont {De~Martini}}, \bibinfo {author} {\bibfnamefont {L.}~\bibnamefont {Hardy}},\ and\ \bibinfo {author} {\bibfnamefont {S.}~\bibnamefont {Popescu}},\ }\bibfield  {title} {\bibinfo {title} {Experimental realization of teleporting an unknown pure quantum state via dual classical and einstein-podolsky-rosen channels},\ }\href@noop {} {\bibfield  {journal} {\bibinfo  {journal} {Phys. Rev. Lett.}\ }\textbf {\bibinfo {volume} {80}},\ \bibinfo {pages} {1121} (\bibinfo {year} {1998})}\BibitemShut {NoStop}%
\bibitem [{\citenamefont {Furusawa}\ \emph {et~al.}(1998)\citenamefont {Furusawa}, \citenamefont {S{\o}rensen}, \citenamefont {Braunstein}, \citenamefont {Fuchs}, \citenamefont {Kimble},\ and\ \citenamefont {Polzik}}]{furusawa1998unconditional}%
  \BibitemOpen
  \bibfield  {author} {\bibinfo {author} {\bibfnamefont {A.}~\bibnamefont {Furusawa}}, \bibinfo {author} {\bibfnamefont {J.~L.}\ \bibnamefont {S{\o}rensen}}, \bibinfo {author} {\bibfnamefont {S.~L.}\ \bibnamefont {Braunstein}}, \bibinfo {author} {\bibfnamefont {C.~A.}\ \bibnamefont {Fuchs}}, \bibinfo {author} {\bibfnamefont {H.~J.}\ \bibnamefont {Kimble}},\ and\ \bibinfo {author} {\bibfnamefont {E.~S.}\ \bibnamefont {Polzik}},\ }\bibfield  {title} {\bibinfo {title} {Unconditional quantum teleportation},\ }\href@noop {} {\bibfield  {journal} {\bibinfo  {journal} {science}\ }\textbf {\bibinfo {volume} {282}},\ \bibinfo {pages} {706} (\bibinfo {year} {1998})}\BibitemShut {NoStop}%
\bibitem [{\citenamefont {Olmschenk}\ \emph {et~al.}(2009)\citenamefont {Olmschenk}, \citenamefont {Matsukevich}, \citenamefont {Maunz}, \citenamefont {Hayes}, \citenamefont {Duan},\ and\ \citenamefont {Monroe}}]{olmschenk2009quantum}%
  \BibitemOpen
  \bibfield  {author} {\bibinfo {author} {\bibfnamefont {S.}~\bibnamefont {Olmschenk}}, \bibinfo {author} {\bibfnamefont {D.}~\bibnamefont {Matsukevich}}, \bibinfo {author} {\bibfnamefont {P.}~\bibnamefont {Maunz}}, \bibinfo {author} {\bibfnamefont {D.}~\bibnamefont {Hayes}}, \bibinfo {author} {\bibfnamefont {L.-M.}\ \bibnamefont {Duan}},\ and\ \bibinfo {author} {\bibfnamefont {C.}~\bibnamefont {Monroe}},\ }\bibfield  {title} {\bibinfo {title} {Quantum teleportation between distant matter qubits},\ }\href@noop {} {\bibfield  {journal} {\bibinfo  {journal} {Science}\ }\textbf {\bibinfo {volume} {323}},\ \bibinfo {pages} {486} (\bibinfo {year} {2009})}\BibitemShut {NoStop}%
\bibitem [{\citenamefont {Yin}\ \emph {et~al.}(2012)\citenamefont {Yin}, \citenamefont {Ren}, \citenamefont {Lu}, \citenamefont {Cao}, \citenamefont {Yong}, \citenamefont {Wu}, \citenamefont {Liu}, \citenamefont {Liao}, \citenamefont {Zhou}, \citenamefont {Jiang} \emph {et~al.}}]{yin2012quantum}%
  \BibitemOpen
  \bibfield  {author} {\bibinfo {author} {\bibfnamefont {J.}~\bibnamefont {Yin}}, \bibinfo {author} {\bibfnamefont {J.-G.}\ \bibnamefont {Ren}}, \bibinfo {author} {\bibfnamefont {H.}~\bibnamefont {Lu}}, \bibinfo {author} {\bibfnamefont {Y.}~\bibnamefont {Cao}}, \bibinfo {author} {\bibfnamefont {H.-L.}\ \bibnamefont {Yong}}, \bibinfo {author} {\bibfnamefont {Y.-P.}\ \bibnamefont {Wu}}, \bibinfo {author} {\bibfnamefont {C.}~\bibnamefont {Liu}}, \bibinfo {author} {\bibfnamefont {S.-K.}\ \bibnamefont {Liao}}, \bibinfo {author} {\bibfnamefont {F.}~\bibnamefont {Zhou}}, \bibinfo {author} {\bibfnamefont {Y.}~\bibnamefont {Jiang}}, \emph {et~al.},\ }\bibfield  {title} {\bibinfo {title} {Quantum teleportation and entanglement distribution over 100-kilometre free-space channels},\ }\href@noop {} {\bibfield  {journal} {\bibinfo  {journal} {Nature}\ }\textbf {\bibinfo {volume} {488}},\ \bibinfo {pages} {185} (\bibinfo {year} {2012})}\BibitemShut {NoStop}%
\bibitem [{\citenamefont {Gottesman}\ and\ \citenamefont {Chuang}(1999)}]{gottesman1999demonstrating}%
  \BibitemOpen
  \bibfield  {author} {\bibinfo {author} {\bibfnamefont {D.}~\bibnamefont {Gottesman}}\ and\ \bibinfo {author} {\bibfnamefont {I.~L.}\ \bibnamefont {Chuang}},\ }\bibfield  {title} {\bibinfo {title} {Demonstrating the viability of universal quantum computation using teleportation and single-qubit operations},\ }\href@noop {} {\bibfield  {journal} {\bibinfo  {journal} {Nature}\ }\textbf {\bibinfo {volume} {402}},\ \bibinfo {pages} {390} (\bibinfo {year} {1999})}\BibitemShut {NoStop}%
\bibitem [{\citenamefont {P\'erez-Delgado}\ and\ \citenamefont {Fitzsimons}(2015)}]{PhysRevLett.114.220502}%
  \BibitemOpen
  \bibfield  {author} {\bibinfo {author} {\bibfnamefont {C.~A.}\ \bibnamefont {P\'erez-Delgado}}\ and\ \bibinfo {author} {\bibfnamefont {J.~F.}\ \bibnamefont {Fitzsimons}},\ }\bibfield  {title} {\bibinfo {title} {Iterated gate teleportation and blind quantum computation},\ }\href@noop {} {\bibfield  {journal} {\bibinfo  {journal} {Phys. Rev. Lett.}\ }\textbf {\bibinfo {volume} {114}},\ \bibinfo {pages} {220502} (\bibinfo {year} {2015})}\BibitemShut {NoStop}%
\bibitem [{\citenamefont {Gisin}\ and\ \citenamefont {Thew}(2007)}]{gisin2007quantum}%
  \BibitemOpen
  \bibfield  {author} {\bibinfo {author} {\bibfnamefont {N.}~\bibnamefont {Gisin}}\ and\ \bibinfo {author} {\bibfnamefont {R.}~\bibnamefont {Thew}},\ }\bibfield  {title} {\bibinfo {title} {Quantum communication},\ }\href@noop {} {\bibfield  {journal} {\bibinfo  {journal} {Nature photonics}\ }\textbf {\bibinfo {volume} {1}},\ \bibinfo {pages} {165} (\bibinfo {year} {2007})}\BibitemShut {NoStop}%
\bibitem [{\citenamefont {Pirandola}\ \emph {et~al.}(2017)\citenamefont {Pirandola}, \citenamefont {Laurenza}, \citenamefont {Ottaviani},\ and\ \citenamefont {Banchi}}]{pirandola2017fundamental}%
  \BibitemOpen
  \bibfield  {author} {\bibinfo {author} {\bibfnamefont {S.}~\bibnamefont {Pirandola}}, \bibinfo {author} {\bibfnamefont {R.}~\bibnamefont {Laurenza}}, \bibinfo {author} {\bibfnamefont {C.}~\bibnamefont {Ottaviani}},\ and\ \bibinfo {author} {\bibfnamefont {L.}~\bibnamefont {Banchi}},\ }\bibfield  {title} {\bibinfo {title} {Fundamental limits of repeaterless quantum communications},\ }\href@noop {} {\bibfield  {journal} {\bibinfo  {journal} {Nature communications}\ }\textbf {\bibinfo {volume} {8}},\ \bibinfo {pages} {15043} (\bibinfo {year} {2017})}\BibitemShut {NoStop}%
\bibitem [{\citenamefont {Huang}\ \emph {et~al.}(2020)\citenamefont {Huang}, \citenamefont {Huang},\ and\ \citenamefont {Li}}]{huang2020identification}%
  \BibitemOpen
  \bibfield  {author} {\bibinfo {author} {\bibfnamefont {N.-N.}\ \bibnamefont {Huang}}, \bibinfo {author} {\bibfnamefont {W.-H.}\ \bibnamefont {Huang}},\ and\ \bibinfo {author} {\bibfnamefont {C.-M.}\ \bibnamefont {Li}},\ }\bibfield  {title} {\bibinfo {title} {Identification of networking quantum teleportation on 14-qubit ibm universal quantum computer},\ }\href@noop {} {\bibfield  {journal} {\bibinfo  {journal} {Scientific reports}\ }\textbf {\bibinfo {volume} {10}},\ \bibinfo {pages} {3093} (\bibinfo {year} {2020})}\BibitemShut {NoStop}%
\bibitem [{\citenamefont {Thomas}\ \emph {et~al.}(2024)\citenamefont {Thomas}, \citenamefont {Yeh}, \citenamefont {Chen}, \citenamefont {Mambretti}, \citenamefont {Kohlert}, \citenamefont {Kanter},\ and\ \citenamefont {Kumar}}]{WOS:001396450700003}%
  \BibitemOpen
  \bibfield  {author} {\bibinfo {author} {\bibfnamefont {J.~m.}\ \bibnamefont {Thomas}}, \bibinfo {author} {\bibfnamefont {F.~i.}\ \bibnamefont {Yeh}}, \bibinfo {author} {\bibfnamefont {J.~h.}\ \bibnamefont {Chen}}, \bibinfo {author} {\bibfnamefont {J.~j.}\ \bibnamefont {Mambretti}}, \bibinfo {author} {\bibfnamefont {S.~j.}\ \bibnamefont {Kohlert}}, \bibinfo {author} {\bibfnamefont {G.~s.}\ \bibnamefont {Kanter}},\ and\ \bibinfo {author} {\bibfnamefont {P.}~\bibnamefont {Kumar}},\ }\bibfield  {title} {\bibinfo {title} {Quantum teleportation coexisting with classical communications in optical fiber},\ }\href@noop {} {\bibfield  {journal} {\bibinfo  {journal} {OPTICA}\ }\textbf {\bibinfo {volume} {11}},\ \bibinfo {pages} {1700} (\bibinfo {year} {2024})}\BibitemShut {NoStop}%
\bibitem [{\citenamefont {Verstraete}\ and\ \citenamefont {Verschelde}(2002)}]{verstraete2002fidelity}%
  \BibitemOpen
  \bibfield  {author} {\bibinfo {author} {\bibfnamefont {F.}~\bibnamefont {Verstraete}}\ and\ \bibinfo {author} {\bibfnamefont {H.}~\bibnamefont {Verschelde}},\ }\bibfield  {title} {\bibinfo {title} {Fidelity of mixed states of two qubits},\ }\href@noop {} {\bibfield  {journal} {\bibinfo  {journal} {Physical Review A}\ }\textbf {\bibinfo {volume} {66}},\ \bibinfo {pages} {022307} (\bibinfo {year} {2002})}\BibitemShut {NoStop}%
\bibitem [{\citenamefont {Jiang}(2019{\natexlab{a}})}]{jiang2019quantum}%
  \BibitemOpen
  \bibfield  {author} {\bibinfo {author} {\bibfnamefont {L.-N.}\ \bibnamefont {Jiang}},\ }\bibfield  {title} {\bibinfo {title} {Quantum teleportation under different collective noise environment},\ }\href@noop {} {\bibfield  {journal} {\bibinfo  {journal} {International Journal of Theoretical Physics}\ }\textbf {\bibinfo {volume} {58}},\ \bibinfo {pages} {522} (\bibinfo {year} {2019}{\natexlab{a}})}\BibitemShut {NoStop}%
\bibitem [{\citenamefont {Horodecki}\ \emph {et~al.}(1996)\citenamefont {Horodecki}, \citenamefont {Horodecki},\ and\ \citenamefont {Horodecki}}]{horodecki1996teleportation}%
  \BibitemOpen
  \bibfield  {author} {\bibinfo {author} {\bibfnamefont {R.}~\bibnamefont {Horodecki}}, \bibinfo {author} {\bibfnamefont {M.}~\bibnamefont {Horodecki}},\ and\ \bibinfo {author} {\bibfnamefont {P.}~\bibnamefont {Horodecki}},\ }\bibfield  {title} {\bibinfo {title} {Teleportation, bell's inequalities and inseparability},\ }\href@noop {} {\bibfield  {journal} {\bibinfo  {journal} {Physics Letters A}\ }\textbf {\bibinfo {volume} {222}},\ \bibinfo {pages} {21} (\bibinfo {year} {1996})}\BibitemShut {NoStop}%
\bibitem [{\citenamefont {Popescu}(1994)}]{popescu1994bell}%
  \BibitemOpen
  \bibfield  {author} {\bibinfo {author} {\bibfnamefont {S.}~\bibnamefont {Popescu}},\ }\bibfield  {title} {\bibinfo {title} {Bell’s inequalities versus teleportation: What is nonlocality?},\ }\href@noop {} {\bibfield  {journal} {\bibinfo  {journal} {Physical review letters}\ }\textbf {\bibinfo {volume} {72}},\ \bibinfo {pages} {797} (\bibinfo {year} {1994})}\BibitemShut {NoStop}%
\bibitem [{\citenamefont {Yu}\ and\ \citenamefont {Eberly}(2004)}]{yu2004finite}%
  \BibitemOpen
  \bibfield  {author} {\bibinfo {author} {\bibfnamefont {T.}~\bibnamefont {Yu}}\ and\ \bibinfo {author} {\bibfnamefont {J.}~\bibnamefont {Eberly}},\ }\bibfield  {title} {\bibinfo {title} {Finite-time disentanglement via spontaneous emission},\ }\href@noop {} {\bibfield  {journal} {\bibinfo  {journal} {Physical Review Letters}\ }\textbf {\bibinfo {volume} {93}},\ \bibinfo {pages} {140404} (\bibinfo {year} {2004})}\BibitemShut {NoStop}%
\bibitem [{\citenamefont {Carvalho}\ \emph {et~al.}(2004)\citenamefont {Carvalho}, \citenamefont {Mintert},\ and\ \citenamefont {Buchleitner}}]{carvalho2004decoherence}%
  \BibitemOpen
  \bibfield  {author} {\bibinfo {author} {\bibfnamefont {A.~R.}\ \bibnamefont {Carvalho}}, \bibinfo {author} {\bibfnamefont {F.}~\bibnamefont {Mintert}},\ and\ \bibinfo {author} {\bibfnamefont {A.}~\bibnamefont {Buchleitner}},\ }\bibfield  {title} {\bibinfo {title} {Decoherence and multipartite entanglement},\ }\href@noop {} {\bibfield  {journal} {\bibinfo  {journal} {Physical review letters}\ }\textbf {\bibinfo {volume} {93}},\ \bibinfo {pages} {230501} (\bibinfo {year} {2004})}\BibitemShut {NoStop}%
\bibitem [{\citenamefont {Harlender}\ and\ \citenamefont {Roszak}(2022)}]{harlender2022transfer}%
  \BibitemOpen
  \bibfield  {author} {\bibinfo {author} {\bibfnamefont {T.}~\bibnamefont {Harlender}}\ and\ \bibinfo {author} {\bibfnamefont {K.}~\bibnamefont {Roszak}},\ }\bibfield  {title} {\bibinfo {title} {Transfer and teleportation of system-environment entanglement},\ }\href@noop {} {\bibfield  {journal} {\bibinfo  {journal} {Physical Review A}\ }\textbf {\bibinfo {volume} {105}},\ \bibinfo {pages} {012407} (\bibinfo {year} {2022})}\BibitemShut {NoStop}%
\bibitem [{\citenamefont {Horodecki}\ \emph {et~al.}(1999)\citenamefont {Horodecki}, \citenamefont {Horodecki},\ and\ \citenamefont {Horodecki}}]{horodecki1999general}%
  \BibitemOpen
  \bibfield  {author} {\bibinfo {author} {\bibfnamefont {M.}~\bibnamefont {Horodecki}}, \bibinfo {author} {\bibfnamefont {P.}~\bibnamefont {Horodecki}},\ and\ \bibinfo {author} {\bibfnamefont {R.}~\bibnamefont {Horodecki}},\ }\bibfield  {title} {\bibinfo {title} {General teleportation channel, singlet fraction, and quasidistillation},\ }\href@noop {} {\bibfield  {journal} {\bibinfo  {journal} {Physical Review A}\ }\textbf {\bibinfo {volume} {60}},\ \bibinfo {pages} {1888} (\bibinfo {year} {1999})}\BibitemShut {NoStop}%
\bibitem [{\citenamefont {Zhang}\ and\ \citenamefont {Wang}(2021)}]{zhang2021entanglement}%
  \BibitemOpen
  \bibfield  {author} {\bibinfo {author} {\bibfnamefont {K.}~\bibnamefont {Zhang}}\ and\ \bibinfo {author} {\bibfnamefont {J.}~\bibnamefont {Wang}},\ }\bibfield  {title} {\bibinfo {title} {Entanglement versus bell nonlocality of quantum nonequilibrium steady states},\ }\href@noop {} {\bibfield  {journal} {\bibinfo  {journal} {Quantum Information Processing}\ }\textbf {\bibinfo {volume} {20}},\ \bibinfo {pages} {147} (\bibinfo {year} {2021})}\BibitemShut {NoStop}%
\bibitem [{\citenamefont {Mazzola}\ \emph {et~al.}(2010)\citenamefont {Mazzola}, \citenamefont {Piilo},\ and\ \citenamefont {Maniscalco}}]{mazzola2010sudden}%
  \BibitemOpen
  \bibfield  {author} {\bibinfo {author} {\bibfnamefont {L.}~\bibnamefont {Mazzola}}, \bibinfo {author} {\bibfnamefont {J.}~\bibnamefont {Piilo}},\ and\ \bibinfo {author} {\bibfnamefont {S.}~\bibnamefont {Maniscalco}},\ }\bibfield  {title} {\bibinfo {title} {Sudden transition between classical and quantum decoherence},\ }\href@noop {} {\bibfield  {journal} {\bibinfo  {journal} {Physical review letters}\ }\textbf {\bibinfo {volume} {104}},\ \bibinfo {pages} {200401} (\bibinfo {year} {2010})}\BibitemShut {NoStop}%
\bibitem [{\citenamefont {Li}\ \emph {et~al.}(2021)\citenamefont {Li}, \citenamefont {Zhu}, \citenamefont {Liang}, \citenamefont {Ye},\ and\ \citenamefont {Fei}}]{li2021quantum}%
  \BibitemOpen
  \bibfield  {author} {\bibinfo {author} {\bibfnamefont {B.}~\bibnamefont {Li}}, \bibinfo {author} {\bibfnamefont {C.-L.}\ \bibnamefont {Zhu}}, \bibinfo {author} {\bibfnamefont {X.-B.}\ \bibnamefont {Liang}}, \bibinfo {author} {\bibfnamefont {B.-L.}\ \bibnamefont {Ye}},\ and\ \bibinfo {author} {\bibfnamefont {S.-M.}\ \bibnamefont {Fei}},\ }\bibfield  {title} {\bibinfo {title} {Quantum discord for multiqubit systems},\ }\href@noop {} {\bibfield  {journal} {\bibinfo  {journal} {Physical Review A}\ }\textbf {\bibinfo {volume} {104}},\ \bibinfo {pages} {012428} (\bibinfo {year} {2021})}\BibitemShut {NoStop}%
\bibitem [{\citenamefont {Radhakrishnan}\ \emph {et~al.}(2020)\citenamefont {Radhakrishnan}, \citenamefont {Lauri{\`e}re},\ and\ \citenamefont {Byrnes}}]{radhakrishnan2020multipartite}%
  \BibitemOpen
  \bibfield  {author} {\bibinfo {author} {\bibfnamefont {C.}~\bibnamefont {Radhakrishnan}}, \bibinfo {author} {\bibfnamefont {M.}~\bibnamefont {Lauri{\`e}re}},\ and\ \bibinfo {author} {\bibfnamefont {T.}~\bibnamefont {Byrnes}},\ }\bibfield  {title} {\bibinfo {title} {Multipartite generalization of quantum discord},\ }\href@noop {} {\bibfield  {journal} {\bibinfo  {journal} {Physical Review Letters}\ }\textbf {\bibinfo {volume} {124}},\ \bibinfo {pages} {110401} (\bibinfo {year} {2020})}\BibitemShut {NoStop}%
\bibitem [{\citenamefont {Badziag}\ \emph {et~al.}(2000)\citenamefont {Badziag}, \citenamefont {Horodecki}, \citenamefont {Horodecki},\ and\ \citenamefont {Horodecki}}]{badziag2000local}%
  \BibitemOpen
  \bibfield  {author} {\bibinfo {author} {\bibfnamefont {P.}~\bibnamefont {Badziag}}, \bibinfo {author} {\bibfnamefont {M.}~\bibnamefont {Horodecki}}, \bibinfo {author} {\bibfnamefont {P.}~\bibnamefont {Horodecki}},\ and\ \bibinfo {author} {\bibfnamefont {R.}~\bibnamefont {Horodecki}},\ }\bibfield  {title} {\bibinfo {title} {Local environment can enhance fidelity of quantum teleportation},\ }\href@noop {} {\bibfield  {journal} {\bibinfo  {journal} {Physical Review A}\ }\textbf {\bibinfo {volume} {62}},\ \bibinfo {pages} {012311} (\bibinfo {year} {2000})}\BibitemShut {NoStop}%
\bibitem [{\citenamefont {Bandyopadhyay}(2002)}]{bandyopadhyay2002origin}%
  \BibitemOpen
  \bibfield  {author} {\bibinfo {author} {\bibfnamefont {S.}~\bibnamefont {Bandyopadhyay}},\ }\bibfield  {title} {\bibinfo {title} {Origin of noisy states whose teleportation fidelity can be enhanced through dissipation},\ }\href@noop {} {\bibfield  {journal} {\bibinfo  {journal} {Physical Review A}\ }\textbf {\bibinfo {volume} {65}},\ \bibinfo {pages} {022302} (\bibinfo {year} {2002})}\BibitemShut {NoStop}%
\bibitem [{\citenamefont {Jiang}(2019{\natexlab{b}})}]{jiang2019quantumlocal}%
  \BibitemOpen
  \bibfield  {author} {\bibinfo {author} {\bibfnamefont {L.-N.}\ \bibnamefont {Jiang}},\ }\bibfield  {title} {\bibinfo {title} {Quantum teleportation under different local independent noise environment},\ }\href@noop {} {\bibfield  {journal} {\bibinfo  {journal} {International Journal of Theoretical Physics}\ }\textbf {\bibinfo {volume} {58}},\ \bibinfo {pages} {3899} (\bibinfo {year} {2019}{\natexlab{b}})}\BibitemShut {NoStop}%
\bibitem [{\citenamefont {Oh}\ \emph {et~al.}(2002)\citenamefont {Oh}, \citenamefont {Lee},\ and\ \citenamefont {Lee}}]{oh2002fidelity}%
  \BibitemOpen
  \bibfield  {author} {\bibinfo {author} {\bibfnamefont {S.}~\bibnamefont {Oh}}, \bibinfo {author} {\bibfnamefont {S.}~\bibnamefont {Lee}},\ and\ \bibinfo {author} {\bibfnamefont {H.-w.}\ \bibnamefont {Lee}},\ }\bibfield  {title} {\bibinfo {title} {Fidelity of quantum teleportation through noisy channels},\ }\href@noop {} {\bibfield  {journal} {\bibinfo  {journal} {Physical Review A}\ }\textbf {\bibinfo {volume} {66}},\ \bibinfo {pages} {022316} (\bibinfo {year} {2002})}\BibitemShut {NoStop}%
\bibitem [{\citenamefont {Xie}(2021)}]{xie2021steady}%
  \BibitemOpen
  \bibfield  {author} {\bibinfo {author} {\bibfnamefont {Y.-X.}\ \bibnamefont {Xie}},\ }\bibfield  {title} {\bibinfo {title} {Steady-state teleportation fidelity and bell nonlocality in dissipative environments},\ }\href@noop {} {\bibfield  {journal} {\bibinfo  {journal} {Communications in Theoretical Physics}\ }\textbf {\bibinfo {volume} {73}},\ \bibinfo {pages} {075102} (\bibinfo {year} {2021})}\BibitemShut {NoStop}%
\bibitem [{\citenamefont {Man}\ and\ \citenamefont {Xia}(2012)}]{man2012quantum}%
  \BibitemOpen
  \bibfield  {author} {\bibinfo {author} {\bibfnamefont {Z.-X.}\ \bibnamefont {Man}}\ and\ \bibinfo {author} {\bibfnamefont {Y.-J.}\ \bibnamefont {Xia}},\ }\bibfield  {title} {\bibinfo {title} {Quantum teleportation in a dissipative environment},\ }\href@noop {} {\bibfield  {journal} {\bibinfo  {journal} {Quantum Information Processing}\ }\textbf {\bibinfo {volume} {11}},\ \bibinfo {pages} {1911} (\bibinfo {year} {2012})}\BibitemShut {NoStop}%
\bibitem [{\citenamefont {Xu}\ and\ \citenamefont {An}(2024)}]{PhysRevA.110.012442}%
  \BibitemOpen
  \bibfield  {author} {\bibinfo {author} {\bibfnamefont {Z.-J.}\ \bibnamefont {Xu}}\ and\ \bibinfo {author} {\bibfnamefont {J.-H.}\ \bibnamefont {An}},\ }\bibfield  {title} {\bibinfo {title} {Noise mitigation in quantum teleportation},\ }\href@noop {} {\bibfield  {journal} {\bibinfo  {journal} {Phys. Rev. A}\ }\textbf {\bibinfo {volume} {110}},\ \bibinfo {pages} {012442} (\bibinfo {year} {2024})}\BibitemShut {NoStop}%
\bibitem [{\citenamefont {Wu}\ and\ \citenamefont {Segal}(2011)}]{wu2011quantum}%
  \BibitemOpen
  \bibfield  {author} {\bibinfo {author} {\bibfnamefont {L.-A.}\ \bibnamefont {Wu}}\ and\ \bibinfo {author} {\bibfnamefont {D.}~\bibnamefont {Segal}},\ }\bibfield  {title} {\bibinfo {title} {Quantum effects in thermal conduction: Nonequilibrium quantum discord and entanglement},\ }\href@noop {} {\bibfield  {journal} {\bibinfo  {journal} {Physical Review A—Atomic, Molecular, and Optical Physics}\ }\textbf {\bibinfo {volume} {84}},\ \bibinfo {pages} {012319} (\bibinfo {year} {2011})}\BibitemShut {NoStop}%
\bibitem [{\citenamefont {Lambert}\ \emph {et~al.}(2007)\citenamefont {Lambert}, \citenamefont {Aguado},\ and\ \citenamefont {Brandes}}]{lambert2007nonequilibrium}%
  \BibitemOpen
  \bibfield  {author} {\bibinfo {author} {\bibfnamefont {N.}~\bibnamefont {Lambert}}, \bibinfo {author} {\bibfnamefont {R.}~\bibnamefont {Aguado}},\ and\ \bibinfo {author} {\bibfnamefont {T.}~\bibnamefont {Brandes}},\ }\bibfield  {title} {\bibinfo {title} {Nonequilibrium entanglement and noise in coupled qubits},\ }\href@noop {} {\bibfield  {journal} {\bibinfo  {journal} {Physical Review B—Condensed Matter and Materials Physics}\ }\textbf {\bibinfo {volume} {75}},\ \bibinfo {pages} {045340} (\bibinfo {year} {2007})}\BibitemShut {NoStop}%
\bibitem [{\citenamefont {Quiroga}\ \emph {et~al.}(2007)\citenamefont {Quiroga}, \citenamefont {Rodriguez}, \citenamefont {Ramirez},\ and\ \citenamefont {Paris}}]{quiroga2007nonequilibrium}%
  \BibitemOpen
  \bibfield  {author} {\bibinfo {author} {\bibfnamefont {L.}~\bibnamefont {Quiroga}}, \bibinfo {author} {\bibfnamefont {F.~J.}\ \bibnamefont {Rodriguez}}, \bibinfo {author} {\bibfnamefont {M.~E.}\ \bibnamefont {Ramirez}},\ and\ \bibinfo {author} {\bibfnamefont {R.}~\bibnamefont {Paris}},\ }\bibfield  {title} {\bibinfo {title} {Nonequilibrium thermal entanglement},\ }\href@noop {} {\bibfield  {journal} {\bibinfo  {journal} {Physical Review A—Atomic, Molecular, and Optical Physics}\ }\textbf {\bibinfo {volume} {75}},\ \bibinfo {pages} {032308} (\bibinfo {year} {2007})}\BibitemShut {NoStop}%
\bibitem [{\citenamefont {Sinaysky}\ \emph {et~al.}(2008{\natexlab{a}})\citenamefont {Sinaysky}, \citenamefont {Petruccione},\ and\ \citenamefont {Burgarth}}]{sinaysky2008dynamics}%
  \BibitemOpen
  \bibfield  {author} {\bibinfo {author} {\bibfnamefont {I.}~\bibnamefont {Sinaysky}}, \bibinfo {author} {\bibfnamefont {F.}~\bibnamefont {Petruccione}},\ and\ \bibinfo {author} {\bibfnamefont {D.}~\bibnamefont {Burgarth}},\ }\bibfield  {title} {\bibinfo {title} {Dynamics of nonequilibrium thermal entanglement},\ }\href@noop {} {\bibfield  {journal} {\bibinfo  {journal} {Physical Review A—Atomic, Molecular, and Optical Physics}\ }\textbf {\bibinfo {volume} {78}},\ \bibinfo {pages} {062301} (\bibinfo {year} {2008}{\natexlab{a}})}\BibitemShut {NoStop}%
\bibitem [{\citenamefont {Wang}\ \emph {et~al.}(2019)\citenamefont {Wang}, \citenamefont {Wu},\ and\ \citenamefont {Wang}}]{wang2018steady}%
  \BibitemOpen
  \bibfield  {author} {\bibinfo {author} {\bibfnamefont {Z.}~\bibnamefont {Wang}}, \bibinfo {author} {\bibfnamefont {W.}~\bibnamefont {Wu}},\ and\ \bibinfo {author} {\bibfnamefont {J.}~\bibnamefont {Wang}},\ }\bibfield  {title} {\bibinfo {title} {Steady-state entanglement and coherence of two coupled qubits in equilibrium and nonequilibrium environments},\ }\href@noop {} {\bibfield  {journal} {\bibinfo  {journal} {Physical Review A}\ }\textbf {\bibinfo {volume} {99}},\ \bibinfo {pages} {042320} (\bibinfo {year} {2019})}\BibitemShut {NoStop}%
\bibitem [{\citenamefont {Kitajima}\ \emph {et~al.}(2014)\citenamefont {Kitajima}, \citenamefont {Arimitsu}, \citenamefont {Obinata},\ and\ \citenamefont {Yoshida}}]{kitajima2014application}%
  \BibitemOpen
  \bibfield  {author} {\bibinfo {author} {\bibfnamefont {S.}~\bibnamefont {Kitajima}}, \bibinfo {author} {\bibfnamefont {T.}~\bibnamefont {Arimitsu}}, \bibinfo {author} {\bibfnamefont {M.}~\bibnamefont {Obinata}},\ and\ \bibinfo {author} {\bibfnamefont {K.}~\bibnamefont {Yoshida}},\ }\bibfield  {title} {\bibinfo {title} {Application of non-equilibrium thermo field dynamics to quantum teleportation under the environment},\ }\href@noop {} {\bibfield  {journal} {\bibinfo  {journal} {Physica A: Statistical Mechanics and its Applications}\ }\textbf {\bibinfo {volume} {404}},\ \bibinfo {pages} {242} (\bibinfo {year} {2014})}\BibitemShut {NoStop}%
\bibitem [{\citenamefont {Bloch}(1957{\natexlab{a}})}]{bloch1957generalized}%
  \BibitemOpen
  \bibfield  {author} {\bibinfo {author} {\bibfnamefont {F.}~\bibnamefont {Bloch}},\ }\bibfield  {title} {\bibinfo {title} {Generalized theory of relaxation},\ }\href@noop {} {\bibfield  {journal} {\bibinfo  {journal} {Physical Review}\ }\textbf {\bibinfo {volume} {105}},\ \bibinfo {pages} {1206} (\bibinfo {year} {1957}{\natexlab{a}})}\BibitemShut {NoStop}%
\bibitem [{\citenamefont {Redfield}(1957)}]{redfield1957theory}%
  \BibitemOpen
  \bibfield  {author} {\bibinfo {author} {\bibfnamefont {A.~G.}\ \bibnamefont {Redfield}},\ }\bibfield  {title} {\bibinfo {title} {On the theory of relaxation processes},\ }\href@noop {} {\bibfield  {journal} {\bibinfo  {journal} {IBM Journal of Research and Development}\ }\textbf {\bibinfo {volume} {1}},\ \bibinfo {pages} {19} (\bibinfo {year} {1957})}\BibitemShut {NoStop}%
\bibitem [{\citenamefont {Redfield}(1996)}]{redfield1996relaxation}%
  \BibitemOpen
  \bibfield  {author} {\bibinfo {author} {\bibfnamefont {A.}~\bibnamefont {Redfield}},\ }\href@noop {} {\bibinfo {title} {Relaxation theory: density matrix formulation encyclopedia of nuclear magnetic resonance ed dm grant and rk harris}} (\bibinfo {year} {1996})\BibitemShut {NoStop}%
\bibitem [{\citenamefont {Pollard}\ \emph {et~al.}(1996)\citenamefont {Pollard}, \citenamefont {Felts},\ and\ \citenamefont {Friesner}}]{pollard1996advances}%
  \BibitemOpen
  \bibfield  {author} {\bibinfo {author} {\bibfnamefont {W.}~\bibnamefont {Pollard}}, \bibinfo {author} {\bibfnamefont {A.}~\bibnamefont {Felts}},\ and\ \bibinfo {author} {\bibfnamefont {R.}~\bibnamefont {Friesner}},\ }\href@noop {} {\bibinfo {title} {Advances in chemical physics vol 93 ed i prigogine and sa rice}} (\bibinfo {year} {1996})\BibitemShut {NoStop}%
\bibitem [{\citenamefont {Ishizaki}\ and\ \citenamefont {Fleming}(2009)}]{ishizaki2009adequacy}%
  \BibitemOpen
  \bibfield  {author} {\bibinfo {author} {\bibfnamefont {A.}~\bibnamefont {Ishizaki}}\ and\ \bibinfo {author} {\bibfnamefont {G.~R.}\ \bibnamefont {Fleming}},\ }\bibfield  {title} {\bibinfo {title} {On the adequacy of the redfield equation and related approaches to the study of quantum dynamics in electronic energy transfer},\ }\href@noop {} {\bibfield  {journal} {\bibinfo  {journal} {The Journal of chemical physics}\ }\textbf {\bibinfo {volume} {130}} (\bibinfo {year} {2009})}\BibitemShut {NoStop}%
\bibitem [{\citenamefont {Lee}\ \emph {et~al.}(2015)\citenamefont {Lee}, \citenamefont {Moix},\ and\ \citenamefont {Cao}}]{lee2015coherent}%
  \BibitemOpen
  \bibfield  {author} {\bibinfo {author} {\bibfnamefont {C.~K.}\ \bibnamefont {Lee}}, \bibinfo {author} {\bibfnamefont {J.}~\bibnamefont {Moix}},\ and\ \bibinfo {author} {\bibfnamefont {J.}~\bibnamefont {Cao}},\ }\bibfield  {title} {\bibinfo {title} {Coherent quantum transport in disordered systems: A unified polaron treatment of hopping and band-like transport},\ }\href@noop {} {\bibfield  {journal} {\bibinfo  {journal} {The Journal of chemical physics}\ }\textbf {\bibinfo {volume} {142}} (\bibinfo {year} {2015})}\BibitemShut {NoStop}%
\bibitem [{\citenamefont {Novoderezhkin}\ \emph {et~al.}(2004)\citenamefont {Novoderezhkin}, \citenamefont {Yakovlev}, \citenamefont {Van~Grondelle},\ and\ \citenamefont {Shuvalov}}]{novoderezhkin2004coherent}%
  \BibitemOpen
  \bibfield  {author} {\bibinfo {author} {\bibfnamefont {V.~I.}\ \bibnamefont {Novoderezhkin}}, \bibinfo {author} {\bibfnamefont {A.~G.}\ \bibnamefont {Yakovlev}}, \bibinfo {author} {\bibfnamefont {R.}~\bibnamefont {Van~Grondelle}},\ and\ \bibinfo {author} {\bibfnamefont {V.~A.}\ \bibnamefont {Shuvalov}},\ }\bibfield  {title} {\bibinfo {title} {Coherent nuclear and electronic dynamics in primary charge separation in photosynthetic reaction centers: a redfield theory approach},\ }\href@noop {} {\bibfield  {journal} {\bibinfo  {journal} {The Journal of Physical Chemistry B}\ }\textbf {\bibinfo {volume} {108}},\ \bibinfo {pages} {7445} (\bibinfo {year} {2004})}\BibitemShut {NoStop}%
\bibitem [{\citenamefont {Jeske}\ \emph {et~al.}(2015{\natexlab{a}})\citenamefont {Jeske}, \citenamefont {Ing}, \citenamefont {Plenio}, \citenamefont {Huelga},\ and\ \citenamefont {Cole}}]{WOS:000349847000005}%
  \BibitemOpen
  \bibfield  {author} {\bibinfo {author} {\bibfnamefont {J.}~\bibnamefont {Jeske}}, \bibinfo {author} {\bibfnamefont {D.~J.}\ \bibnamefont {Ing}}, \bibinfo {author} {\bibfnamefont {M.~B.}\ \bibnamefont {Plenio}}, \bibinfo {author} {\bibfnamefont {S.~F.}\ \bibnamefont {Huelga}},\ and\ \bibinfo {author} {\bibfnamefont {J.~H.}\ \bibnamefont {Cole}},\ }\bibfield  {title} {\bibinfo {title} {Bloch-redfield equations for modeling light-harvesting complexes},\ }\href@noop {} {\bibfield  {journal} {\bibinfo  {journal} {JOURNAL OF CHEMICAL PHYSICS}\ }\textbf {\bibinfo {volume} {142}} (\bibinfo {year} {2015}{\natexlab{a}})}\BibitemShut {NoStop}%
\bibitem [{\citenamefont {Zhang}\ and\ \citenamefont {Wang}(2014)}]{zhang2014curl}%
  \BibitemOpen
  \bibfield  {author} {\bibinfo {author} {\bibfnamefont {Z.}~\bibnamefont {Zhang}}\ and\ \bibinfo {author} {\bibfnamefont {J.}~\bibnamefont {Wang}},\ }\bibfield  {title} {\bibinfo {title} {Curl flux, coherence, and population landscape of molecular systems: Nonequilibrium quantum steady state, energy (charge) transport, and thermodynamics},\ }\href@noop {} {\bibfield  {journal} {\bibinfo  {journal} {The Journal of chemical physics}\ }\textbf {\bibinfo {volume} {140}} (\bibinfo {year} {2014})}\BibitemShut {NoStop}%
\bibitem [{\citenamefont {Li}\ \emph {et~al.}(2015)\citenamefont {Li}, \citenamefont {Cai},\ and\ \citenamefont {Sun}}]{li2015steady}%
  \BibitemOpen
  \bibfield  {author} {\bibinfo {author} {\bibfnamefont {S.-W.}\ \bibnamefont {Li}}, \bibinfo {author} {\bibfnamefont {C.}~\bibnamefont {Cai}},\ and\ \bibinfo {author} {\bibfnamefont {C.}~\bibnamefont {Sun}},\ }\bibfield  {title} {\bibinfo {title} {Steady quantum coherence in non-equilibrium environment},\ }\href@noop {} {\bibfield  {journal} {\bibinfo  {journal} {Annals of Physics}\ }\textbf {\bibinfo {volume} {360}},\ \bibinfo {pages} {19} (\bibinfo {year} {2015})}\BibitemShut {NoStop}%
\bibitem [{\citenamefont {Huangfu}\ and\ \citenamefont {Jing}(2018)}]{huangfu2018steady}%
  \BibitemOpen
  \bibfield  {author} {\bibinfo {author} {\bibfnamefont {Y.}~\bibnamefont {Huangfu}}\ and\ \bibinfo {author} {\bibfnamefont {J.}~\bibnamefont {Jing}},\ }\bibfield  {title} {\bibinfo {title} {Steady bipartite coherence induced by non-equilibrium environment},\ }\href@noop {} {\bibfield  {journal} {\bibinfo  {journal} {Science China Physics, Mechanics \& Astronomy}\ }\textbf {\bibinfo {volume} {61}},\ \bibinfo {pages} {1} (\bibinfo {year} {2018})}\BibitemShut {NoStop}%
\bibitem [{\citenamefont {Zhang}\ and\ \citenamefont {Wang}(2015{\natexlab{a}})}]{zhang2015landscape}%
  \BibitemOpen
  \bibfield  {author} {\bibinfo {author} {\bibfnamefont {Z.}~\bibnamefont {Zhang}}\ and\ \bibinfo {author} {\bibfnamefont {J.}~\bibnamefont {Wang}},\ }\bibfield  {title} {\bibinfo {title} {Landscape, kinetics, paths and statistics of curl flux, coherence, entanglement and energy transfer in non-equilibrium quantum systems},\ }\href@noop {} {\bibfield  {journal} {\bibinfo  {journal} {New Journal of Physics}\ }\textbf {\bibinfo {volume} {17}},\ \bibinfo {pages} {043053} (\bibinfo {year} {2015}{\natexlab{a}})}\BibitemShut {NoStop}%
\bibitem [{\citenamefont {Zhang}\ and\ \citenamefont {Wang}(2015{\natexlab{b}})}]{zhang2015shape}%
  \BibitemOpen
  \bibfield  {author} {\bibinfo {author} {\bibfnamefont {Z.}~\bibnamefont {Zhang}}\ and\ \bibinfo {author} {\bibfnamefont {J.}~\bibnamefont {Wang}},\ }\bibfield  {title} {\bibinfo {title} {Shape, orientation and magnitude of the curl quantum flux, the coherence and the statistical correlations in energy transport at nonequilibrium steady state},\ }\href@noop {} {\bibfield  {journal} {\bibinfo  {journal} {New Journal of Physics}\ }\textbf {\bibinfo {volume} {17}},\ \bibinfo {pages} {093021} (\bibinfo {year} {2015}{\natexlab{b}})}\BibitemShut {NoStop}%
\bibitem [{\citenamefont {Wang}\ \emph {et~al.}(2018)\citenamefont {Wang}, \citenamefont {Wu}, \citenamefont {Cui},\ and\ \citenamefont {Wang}}]{wang2018coherence}%
  \BibitemOpen
  \bibfield  {author} {\bibinfo {author} {\bibfnamefont {Z.}~\bibnamefont {Wang}}, \bibinfo {author} {\bibfnamefont {W.}~\bibnamefont {Wu}}, \bibinfo {author} {\bibfnamefont {G.}~\bibnamefont {Cui}},\ and\ \bibinfo {author} {\bibfnamefont {J.}~\bibnamefont {Wang}},\ }\bibfield  {title} {\bibinfo {title} {Coherence enhanced quantum metrology in a nonequilibrium optical molecule},\ }\href@noop {} {\bibfield  {journal} {\bibinfo  {journal} {New Journal of Physics}\ }\textbf {\bibinfo {volume} {20}},\ \bibinfo {pages} {033034} (\bibinfo {year} {2018})}\BibitemShut {NoStop}%
\bibitem [{\citenamefont {Guarnieri}\ \emph {et~al.}(2018)\citenamefont {Guarnieri}, \citenamefont {Kol{\'a}{\v{r}}},\ and\ \citenamefont {Filip}}]{guarnieri2018steady}%
  \BibitemOpen
  \bibfield  {author} {\bibinfo {author} {\bibfnamefont {G.}~\bibnamefont {Guarnieri}}, \bibinfo {author} {\bibfnamefont {M.}~\bibnamefont {Kol{\'a}{\v{r}}}},\ and\ \bibinfo {author} {\bibfnamefont {R.}~\bibnamefont {Filip}},\ }\bibfield  {title} {\bibinfo {title} {Steady-state coherences by composite system-bath interactions},\ }\href@noop {} {\bibfield  {journal} {\bibinfo  {journal} {Physical review letters}\ }\textbf {\bibinfo {volume} {121}},\ \bibinfo {pages} {070401} (\bibinfo {year} {2018})}\BibitemShut {NoStop}%
\bibitem [{\citenamefont {Jeske}\ \emph {et~al.}(2015{\natexlab{b}})\citenamefont {Jeske}, \citenamefont {David}, \citenamefont {Plenio}, \citenamefont {Huelga},\ and\ \citenamefont {Cole}}]{jeske2015bloch}%
  \BibitemOpen
  \bibfield  {author} {\bibinfo {author} {\bibfnamefont {J.}~\bibnamefont {Jeske}}, \bibinfo {author} {\bibfnamefont {J.}~\bibnamefont {David}}, \bibinfo {author} {\bibfnamefont {M.~B.}\ \bibnamefont {Plenio}}, \bibinfo {author} {\bibfnamefont {S.~F.}\ \bibnamefont {Huelga}},\ and\ \bibinfo {author} {\bibfnamefont {J.~H.}\ \bibnamefont {Cole}},\ }\bibfield  {title} {\bibinfo {title} {Bloch-redfield equations for modeling light-harvesting complexes},\ }\href@noop {} {\bibfield  {journal} {\bibinfo  {journal} {The Journal of chemical physics}\ }\textbf {\bibinfo {volume} {142}} (\bibinfo {year} {2015}{\natexlab{b}})}\BibitemShut {NoStop}%
\bibitem [{\citenamefont {Spohn}(1980)}]{spohn1980kinetic}%
  \BibitemOpen
  \bibfield  {author} {\bibinfo {author} {\bibfnamefont {H.}~\bibnamefont {Spohn}},\ }\bibfield  {title} {\bibinfo {title} {Kinetic equations from hamiltonian dynamics: Markovian limits},\ }\href@noop {} {\bibfield  {journal} {\bibinfo  {journal} {Reviews of Modern Physics}\ }\textbf {\bibinfo {volume} {52}},\ \bibinfo {pages} {569} (\bibinfo {year} {1980})}\BibitemShut {NoStop}%
\bibitem [{\citenamefont {Su{\'a}rez}\ \emph {et~al.}(1992)\citenamefont {Su{\'a}rez}, \citenamefont {Silbey},\ and\ \citenamefont {Oppenheim}}]{suarez1992memory}%
  \BibitemOpen
  \bibfield  {author} {\bibinfo {author} {\bibfnamefont {A.}~\bibnamefont {Su{\'a}rez}}, \bibinfo {author} {\bibfnamefont {R.}~\bibnamefont {Silbey}},\ and\ \bibinfo {author} {\bibfnamefont {I.}~\bibnamefont {Oppenheim}},\ }\bibfield  {title} {\bibinfo {title} {Memory effects in the relaxation of quantum open systems},\ }\href@noop {} {\bibfield  {journal} {\bibinfo  {journal} {The Journal of chemical physics}\ }\textbf {\bibinfo {volume} {97}},\ \bibinfo {pages} {5101} (\bibinfo {year} {1992})}\BibitemShut {NoStop}%
\bibitem [{\citenamefont {Petrosyan}\ and\ \citenamefont {Kurizki}(2002)}]{PhysRevLett.89.207902}%
  \BibitemOpen
  \bibfield  {author} {\bibinfo {author} {\bibfnamefont {D.}~\bibnamefont {Petrosyan}}\ and\ \bibinfo {author} {\bibfnamefont {G.}~\bibnamefont {Kurizki}},\ }\bibfield  {title} {\bibinfo {title} {Scalable solid-state quantum processor using subradiant two-atom states},\ }\href@noop {} {\bibfield  {journal} {\bibinfo  {journal} {Phys. Rev. Lett.}\ }\textbf {\bibinfo {volume} {89}},\ \bibinfo {pages} {207902} (\bibinfo {year} {2002})}\BibitemShut {NoStop}%
\bibitem [{\citenamefont {Imamo{\=g}lu}\ \emph {et~al.}(1999)\citenamefont {Imamo{\=g}lu}, \citenamefont {Awschalom}, \citenamefont {Burkard}, \citenamefont {DiVincenzo}, \citenamefont {Loss}, \citenamefont {Sherwin},\ and\ \citenamefont {Small}}]{PhysRevLett.83.4204}%
  \BibitemOpen
  \bibfield  {author} {\bibinfo {author} {\bibfnamefont {A.}~\bibnamefont {Imamo{\=g}lu}}, \bibinfo {author} {\bibfnamefont {D.~D.}\ \bibnamefont {Awschalom}}, \bibinfo {author} {\bibfnamefont {G.}~\bibnamefont {Burkard}}, \bibinfo {author} {\bibfnamefont {D.~P.}\ \bibnamefont {DiVincenzo}}, \bibinfo {author} {\bibfnamefont {D.}~\bibnamefont {Loss}}, \bibinfo {author} {\bibfnamefont {M.}~\bibnamefont {Sherwin}},\ and\ \bibinfo {author} {\bibfnamefont {A.}~\bibnamefont {Small}},\ }\bibfield  {title} {\bibinfo {title} {Quantum information processing using quantum dot spins and cavity qed},\ }\href@noop {} {\bibfield  {journal} {\bibinfo  {journal} {Phys. Rev. Lett.}\ }\textbf {\bibinfo {volume} {83}},\ \bibinfo {pages} {4204} (\bibinfo {year} {1999})}\BibitemShut {NoStop}%
\bibitem [{\citenamefont {Liao}\ \emph {et~al.}(2011)\citenamefont {Liao}, \citenamefont {Huang},\ and\ \citenamefont {Kuang}}]{PhysRevA.83.052110}%
  \BibitemOpen
  \bibfield  {author} {\bibinfo {author} {\bibfnamefont {J.-Q.}\ \bibnamefont {Liao}}, \bibinfo {author} {\bibfnamefont {J.-F.}\ \bibnamefont {Huang}},\ and\ \bibinfo {author} {\bibfnamefont {L.-M.}\ \bibnamefont {Kuang}},\ }\bibfield  {title} {\bibinfo {title} {Quantum thermalization of two coupled two-level systems in eigenstate and bare-state representations},\ }\href@noop {} {\bibfield  {journal} {\bibinfo  {journal} {Phys. Rev. A}\ }\textbf {\bibinfo {volume} {83}},\ \bibinfo {pages} {052110} (\bibinfo {year} {2011})}\BibitemShut {NoStop}%
\bibitem [{\citenamefont {Bloch}(1957{\natexlab{b}})}]{PhysRev.105.1206}%
  \BibitemOpen
  \bibfield  {author} {\bibinfo {author} {\bibfnamefont {F.}~\bibnamefont {Bloch}},\ }\bibfield  {title} {\bibinfo {title} {Generalized theory of relaxation},\ }\href@noop {} {\bibfield  {journal} {\bibinfo  {journal} {Phys. Rev.}\ }\textbf {\bibinfo {volume} {105}},\ \bibinfo {pages} {1206} (\bibinfo {year} {1957}{\natexlab{b}})}\BibitemShut {NoStop}%
\bibitem [{\citenamefont {Wang}\ and\ \citenamefont {Wang}(2019)}]{wang2019nonequilibrium}%
  \BibitemOpen
  \bibfield  {author} {\bibinfo {author} {\bibfnamefont {X.}~\bibnamefont {Wang}}\ and\ \bibinfo {author} {\bibfnamefont {J.}~\bibnamefont {Wang}},\ }\bibfield  {title} {\bibinfo {title} {Nonequilibrium effects on quantum correlations: Discord, mutual information, and entanglement of a two-fermionic system in bosonic and fermionic environments},\ }\href@noop {} {\bibfield  {journal} {\bibinfo  {journal} {Physical Review A}\ }\textbf {\bibinfo {volume} {100}},\ \bibinfo {pages} {052331} (\bibinfo {year} {2019})}\BibitemShut {NoStop}%
\bibitem [{\citenamefont {Sinaysky}\ \emph {et~al.}(2008{\natexlab{b}})\citenamefont {Sinaysky}, \citenamefont {Petruccione},\ and\ \citenamefont {Burgarth}}]{PhysRevA.78.062301}%
  \BibitemOpen
  \bibfield  {author} {\bibinfo {author} {\bibfnamefont {I.}~\bibnamefont {Sinaysky}}, \bibinfo {author} {\bibfnamefont {F.}~\bibnamefont {Petruccione}},\ and\ \bibinfo {author} {\bibfnamefont {D.}~\bibnamefont {Burgarth}},\ }\bibfield  {title} {\bibinfo {title} {Dynamics of nonequilibrium thermal entanglement},\ }\href@noop {} {\bibfield  {journal} {\bibinfo  {journal} {Phys. Rev. A}\ }\textbf {\bibinfo {volume} {78}},\ \bibinfo {pages} {062301} (\bibinfo {year} {2008}{\natexlab{b}})}\BibitemShut {NoStop}%
\bibitem [{\citenamefont {Hu}\ \emph {et~al.}(2018)\citenamefont {Hu}, \citenamefont {Man},\ and\ \citenamefont {Xia}}]{hu2018steady}%
  \BibitemOpen
  \bibfield  {author} {\bibinfo {author} {\bibfnamefont {L.-Z.}\ \bibnamefont {Hu}}, \bibinfo {author} {\bibfnamefont {Z.-X.}\ \bibnamefont {Man}},\ and\ \bibinfo {author} {\bibfnamefont {Y.-J.}\ \bibnamefont {Xia}},\ }\bibfield  {title} {\bibinfo {title} {Steady-state entanglement and thermalization of coupled qubits in two common heat baths},\ }\href@noop {} {\bibfield  {journal} {\bibinfo  {journal} {Quantum Information Processing}\ }\textbf {\bibinfo {volume} {17}},\ \bibinfo {pages} {1} (\bibinfo {year} {2018})}\BibitemShut {NoStop}%
\bibitem [{\citenamefont {Nandi}\ \emph {et~al.}(2018)\citenamefont {Nandi}, \citenamefont {Datta}, \citenamefont {Das},\ and\ \citenamefont {Agrawal}}]{nandi2018two}%
  \BibitemOpen
  \bibfield  {author} {\bibinfo {author} {\bibfnamefont {S.}~\bibnamefont {Nandi}}, \bibinfo {author} {\bibfnamefont {C.}~\bibnamefont {Datta}}, \bibinfo {author} {\bibfnamefont {A.}~\bibnamefont {Das}},\ and\ \bibinfo {author} {\bibfnamefont {P.}~\bibnamefont {Agrawal}},\ }\bibfield  {title} {\bibinfo {title} {Two-qubit mixed states and teleportation fidelity: purity, concurrence, and beyond},\ }\href@noop {} {\bibfield  {journal} {\bibinfo  {journal} {The European Physical Journal D}\ }\textbf {\bibinfo {volume} {72}},\ \bibinfo {pages} {1} (\bibinfo {year} {2018})}\BibitemShut {NoStop}%
\bibitem [{\citenamefont {Oosterkamp}\ \emph {et~al.}(1998)\citenamefont {Oosterkamp}, \citenamefont {Fujisawa}, \citenamefont {Van Der~Wiel}, \citenamefont {Ishibashi}, \citenamefont {Hijman}, \citenamefont {Tarucha},\ and\ \citenamefont {Kouwenhoven}}]{oosterkamp1998microwave}%
  \BibitemOpen
  \bibfield  {author} {\bibinfo {author} {\bibfnamefont {T.}~\bibnamefont {Oosterkamp}}, \bibinfo {author} {\bibfnamefont {T.}~\bibnamefont {Fujisawa}}, \bibinfo {author} {\bibfnamefont {W.}~\bibnamefont {Van Der~Wiel}}, \bibinfo {author} {\bibfnamefont {K.}~\bibnamefont {Ishibashi}}, \bibinfo {author} {\bibfnamefont {R.}~\bibnamefont {Hijman}}, \bibinfo {author} {\bibfnamefont {S.}~\bibnamefont {Tarucha}},\ and\ \bibinfo {author} {\bibfnamefont {L.~P.}\ \bibnamefont {Kouwenhoven}},\ }\bibfield  {title} {\bibinfo {title} {Microwave spectroscopy of a quantum-dot molecule},\ }\href@noop {} {\bibfield  {journal} {\bibinfo  {journal} {Nature}\ }\textbf {\bibinfo {volume} {395}},\ \bibinfo {pages} {873} (\bibinfo {year} {1998})}\BibitemShut {NoStop}%
\bibitem [{\citenamefont {Li}\ \emph {et~al.}(2022)\citenamefont {Li}, \citenamefont {Cao}, \citenamefont {Li}, \citenamefont {Cai}, \citenamefont {Liu}, \citenamefont {Ren}, \citenamefont {Liao}, \citenamefont {Wu}, \citenamefont {Li}, \citenamefont {Li}, \citenamefont {Liu}, \citenamefont {Lu}, \citenamefont {Yin}, \citenamefont {Chen}, \citenamefont {Peng},\ and\ \citenamefont {Pan}}]{PhysRevLett.128.170501}%
  \BibitemOpen
  \bibfield  {author} {\bibinfo {author} {\bibfnamefont {B.}~\bibnamefont {Li}}, \bibinfo {author} {\bibfnamefont {Y.}~\bibnamefont {Cao}}, \bibinfo {author} {\bibfnamefont {Y.-H.}\ \bibnamefont {Li}}, \bibinfo {author} {\bibfnamefont {W.-Q.}\ \bibnamefont {Cai}}, \bibinfo {author} {\bibfnamefont {W.-Y.}\ \bibnamefont {Liu}}, \bibinfo {author} {\bibfnamefont {J.-G.}\ \bibnamefont {Ren}}, \bibinfo {author} {\bibfnamefont {S.-K.}\ \bibnamefont {Liao}}, \bibinfo {author} {\bibfnamefont {H.-N.}\ \bibnamefont {Wu}}, \bibinfo {author} {\bibfnamefont {S.-L.}\ \bibnamefont {Li}}, \bibinfo {author} {\bibfnamefont {L.}~\bibnamefont {Li}}, \bibinfo {author} {\bibfnamefont {N.-L.}\ \bibnamefont {Liu}}, \bibinfo {author} {\bibfnamefont {C.-Y.}\ \bibnamefont {Lu}}, \bibinfo {author} {\bibfnamefont {J.}~\bibnamefont {Yin}}, \bibinfo {author} {\bibfnamefont {Y.-A.}\ \bibnamefont {Chen}}, \bibinfo {author} {\bibfnamefont {C.-Z.}\ \bibnamefont {Peng}},\ and\ \bibinfo {author} {\bibfnamefont {J.-W.}\ \bibnamefont {Pan}},\
  }\bibfield  {title} {\bibinfo {title} {Quantum state transfer over 1200 km assisted by prior distributed entanglement},\ }\href@noop {} {\bibfield  {journal} {\bibinfo  {journal} {Phys. Rev. Lett.}\ }\textbf {\bibinfo {volume} {128}},\ \bibinfo {pages} {170501} (\bibinfo {year} {2022})}\BibitemShut {NoStop}%
\bibitem [{\citenamefont {Xia}\ \emph {et~al.}(2017)\citenamefont {Xia}, \citenamefont {Sun}, \citenamefont {Zhang},\ and\ \citenamefont {Pan}}]{xia2017long}%
  \BibitemOpen
  \bibfield  {author} {\bibinfo {author} {\bibfnamefont {X.-X.}\ \bibnamefont {Xia}}, \bibinfo {author} {\bibfnamefont {Q.-C.}\ \bibnamefont {Sun}}, \bibinfo {author} {\bibfnamefont {Q.}~\bibnamefont {Zhang}},\ and\ \bibinfo {author} {\bibfnamefont {J.-W.}\ \bibnamefont {Pan}},\ }\bibfield  {title} {\bibinfo {title} {Long distance quantum teleportation},\ }\href@noop {} {\bibfield  {journal} {\bibinfo  {journal} {Quantum Science and Technology}\ }\textbf {\bibinfo {volume} {3}},\ \bibinfo {pages} {014012} (\bibinfo {year} {2017})}\BibitemShut {NoStop}%
\bibitem [{\citenamefont {Bowen}\ \emph {et~al.}(2003)\citenamefont {Bowen}, \citenamefont {Treps}, \citenamefont {Buchler}, \citenamefont {Schnabel}, \citenamefont {Ralph}, \citenamefont {Bachor}, \citenamefont {Symul},\ and\ \citenamefont {Lam}}]{PhysRevA.67.032302}%
  \BibitemOpen
  \bibfield  {author} {\bibinfo {author} {\bibfnamefont {W.~P.}\ \bibnamefont {Bowen}}, \bibinfo {author} {\bibfnamefont {N.}~\bibnamefont {Treps}}, \bibinfo {author} {\bibfnamefont {B.~C.}\ \bibnamefont {Buchler}}, \bibinfo {author} {\bibfnamefont {R.}~\bibnamefont {Schnabel}}, \bibinfo {author} {\bibfnamefont {T.~C.}\ \bibnamefont {Ralph}}, \bibinfo {author} {\bibfnamefont {H.-A.}\ \bibnamefont {Bachor}}, \bibinfo {author} {\bibfnamefont {T.}~\bibnamefont {Symul}},\ and\ \bibinfo {author} {\bibfnamefont {P.~K.}\ \bibnamefont {Lam}},\ }\bibfield  {title} {\bibinfo {title} {Experimental investigation of continuous-variable quantum teleportation},\ }\href@noop {} {\bibfield  {journal} {\bibinfo  {journal} {Phys. Rev. A}\ }\textbf {\bibinfo {volume} {67}},\ \bibinfo {pages} {032302} (\bibinfo {year} {2003})}\BibitemShut {NoStop}%
\bibitem [{\citenamefont {Pirandola}\ \emph {et~al.}(2021)\citenamefont {Pirandola}, \citenamefont {Ottaviani}, \citenamefont {Jacobsen}, \citenamefont {Spedalieri}, \citenamefont {Braunstein}, \citenamefont {Gehring},\ and\ \citenamefont {Andersen}}]{pirandola2021environment}%
  \BibitemOpen
  \bibfield  {author} {\bibinfo {author} {\bibfnamefont {S.}~\bibnamefont {Pirandola}}, \bibinfo {author} {\bibfnamefont {C.}~\bibnamefont {Ottaviani}}, \bibinfo {author} {\bibfnamefont {C.~S.}\ \bibnamefont {Jacobsen}}, \bibinfo {author} {\bibfnamefont {G.}~\bibnamefont {Spedalieri}}, \bibinfo {author} {\bibfnamefont {S.~L.}\ \bibnamefont {Braunstein}}, \bibinfo {author} {\bibfnamefont {T.}~\bibnamefont {Gehring}},\ and\ \bibinfo {author} {\bibfnamefont {U.~L.}\ \bibnamefont {Andersen}},\ }\bibfield  {title} {\bibinfo {title} {Environment-assisted bosonic quantum communications},\ }\href@noop {} {\bibfield  {journal} {\bibinfo  {journal} {npj Quantum Information}\ }\textbf {\bibinfo {volume} {7}},\ \bibinfo {pages} {77} (\bibinfo {year} {2021})}\BibitemShut {NoStop}%
\bibitem [{\citenamefont {Sun}\ \emph {et~al.}(2016)\citenamefont {Sun}, \citenamefont {Mao}, \citenamefont {Chen}, \citenamefont {Zhang}, \citenamefont {Jiang}, \citenamefont {Zhang}, \citenamefont {Zhang}, \citenamefont {Miki}, \citenamefont {Yamashita}, \citenamefont {Terai} \emph {et~al.}}]{sun2016quantum}%
  \BibitemOpen
  \bibfield  {author} {\bibinfo {author} {\bibfnamefont {Q.-C.}\ \bibnamefont {Sun}}, \bibinfo {author} {\bibfnamefont {Y.-L.}\ \bibnamefont {Mao}}, \bibinfo {author} {\bibfnamefont {S.-J.}\ \bibnamefont {Chen}}, \bibinfo {author} {\bibfnamefont {W.}~\bibnamefont {Zhang}}, \bibinfo {author} {\bibfnamefont {Y.-F.}\ \bibnamefont {Jiang}}, \bibinfo {author} {\bibfnamefont {Y.-B.}\ \bibnamefont {Zhang}}, \bibinfo {author} {\bibfnamefont {W.-J.}\ \bibnamefont {Zhang}}, \bibinfo {author} {\bibfnamefont {S.}~\bibnamefont {Miki}}, \bibinfo {author} {\bibfnamefont {T.}~\bibnamefont {Yamashita}}, \bibinfo {author} {\bibfnamefont {H.}~\bibnamefont {Terai}}, \emph {et~al.},\ }\bibfield  {title} {\bibinfo {title} {Quantum teleportation with independent sources and prior entanglement distribution over a network},\ }\href@noop {} {\bibfield  {journal} {\bibinfo  {journal} {Nature Photonics}\ }\textbf {\bibinfo {volume} {10}},\ \bibinfo {pages} {671} (\bibinfo {year} {2016})}\BibitemShut {NoStop}%
\bibitem [{\citenamefont {Hill}\ and\ \citenamefont {Wootters}(1997)}]{PhysRevLett.78.5022}%
  \BibitemOpen
  \bibfield  {author} {\bibinfo {author} {\bibfnamefont {S.~A.}\ \bibnamefont {Hill}}\ and\ \bibinfo {author} {\bibfnamefont {W.~K.}\ \bibnamefont {Wootters}},\ }\bibfield  {title} {\bibinfo {title} {Entanglement of a pair of quantum bits},\ }\href@noop {} {\bibfield  {journal} {\bibinfo  {journal} {Phys. Rev. Lett.}\ }\textbf {\bibinfo {volume} {78}},\ \bibinfo {pages} {5022} (\bibinfo {year} {1997})}\BibitemShut {NoStop}%
\end{thebibliography}

\providecommand{\noopsort}[1]{}\providecommand{\singleletter}[1]{#1}%

\end{document}